\documentclass[letterpaper,9pt,preprint]{elsarticle}

 \journal{TexExchange}
 
    \journal{Acta Materialia}
    \usepackage{fullpage}
    
    \usepackage{tgtermes}

    \usepackage{hhline}  
    \usepackage{float}
    \usepackage{subfig}
    \usepackage{multirow}
    \usepackage{graphicx,booktabs}
    \usepackage{xfrac}
    \usepackage{titlesec}
    \usepackage{wrapfig,booktabs}
    \usepackage{mwe}
    \usepackage{adjustbox}
    \usepackage{caption}
    \usepackage{makecell}
    
    \usepackage{url}
    \usepackage{amsmath,array}
    \usepackage[usenames, dvipsnames]{color}
    \usepackage{xcolor}
    \usepackage{array}
    \usepackage{ragged2e}
    \usepackage{mathtools}

    \usepackage{MnSymbol}
    \usepackage{xcolor}

    \usepackage{chngcntr}
    \usepackage{setspace}

    \usepackage{tikz}
    \usepackage{color, colortbl}
    \usepackage{multirow}
    
\usetikzlibrary{matrix, shapes,arrows.meta}
\tikzset{
    block/.style={
        rectangle, draw, fill=blue!20, 
        text width=5em, text centered, rounded corners, minimum height=4em
    },
    cloud/.style={
        draw, ellipse,fill=red!20, node distance=3cm,
        minimum height=2em
    },  
    decision/.style={
        diamond, draw, fill=blue!20, 
        text width=4.5em, text badly centered, node distance=3cm, inner sep=0pt
    },
    line/.style = {draw, -Latex},
}

    \usetikzlibrary{plotmarks}
    
    \newcommand{\etal}{\textit{et al}.}

    \fboxsep=\tabcolsep\relax
    \fboxrule=\arrayrulewidth\relax

    \newcolumntype{A}[2]{%
    >{\minipage{\dimexpr#1\linewidth-2\tabcolsep-#2\arrayrulewidth\relax}\vspace\tabcolsep}%
    c<{\vspace\tabcolsep\endminipage}}

    \graphicspath{{images/}}
     
    \usepackage{lineno,hyperref}
    \modulolinenumbers[5]

    \usepackage{cleveref}

    \crefformat{section}{\S#2#1#3} 
    \crefformat{subsection}{\S#2#1#3}
    \crefformat{subsubsection}{\S#2#1#3}
    \biboptions{numbers,sort&compress}

\begin{document}

\begin{frontmatter}


\title{Towards Stacking Fault Energy Engineering in FCC High Entropy Alloys}

\author{T. Khan$^{a,h}$, T. Kirk$^b$, G. Vazquez$^c$, P. Singh$^d$, A.V. Smirnov,$^d$ Duane D. Johnson$^{d,e}$, K. Youssef$^f$, R. Arr\'{o}yave$^{b,c,g}$}

\address{$^a$ Department of Chemical Biology, University of California, Berkeley, Berkeley, CA}
\address{$^h$ Department of Chemistry, Columbia University, New York, NY 10027, USA}
\address{$^b$Department of Mechanical Engineering, Texas A\&M University, College Station, TX 77843, USA}
\address{$^c$Department of Materials Science and Engineering, Texas A\&M University, College Station, TX 77843, USA}
\address{$^d$Ames Laboratory, U.S. Department of Energy, Iowa State University, Ames, Iowa 50011 USA}
\address{$^e$Department of Materials Science \& Engineering, Iowa State University, Ames, Iowa 50011 USA}
\address{$^f$Qatar University, Doha, Qatar}
\address{$^g$Department of Industrial and Systems Engineering, Texas A\&M University, College Station, TX 77843, USA}
\cortext[mycorrespondingauthor]{Corresponding author email:\textrm{rarroyave@tamu.edu}} 

\begin{abstract}
Stacking Fault Energy (SFE) is an intrinsic alloy property that governs much of the plastic deformation mechanisms observed in fcc alloys. While SFE has been recognized for many years as a key intrinsic mechanical property, its inference via experimental observations or prediction using, for example, computationally intensive first-principles methods is challenging. {This difficulty precludes the explicit use of SFE as an alloy design parameter}. In this work, we combine DFT calculations (with necessary configurational averaging), machine-learning (ML) and physics-based models to predict the SFE in the fcc CoCrFeMnNiV-Al high-entropy alloy space. The best-performing ML  model is capable of accurately predicting the SFE of arbitrary compositions within this 7-element system. This efficient model along with a recently developed model to estimate  intrinsic strength of fcc HEAs is used to explore the strength--SFE Pareto front,  predicting  new-candidate alloys {with particularly} interesting mechanical behavior.
\end{abstract}


\end{frontmatter}

\section{Introduction}
\label{sec:intro}

High Entropy Alloys (HEAs) {differ from their conventional counterparts in that they tend to be located within the center of the composition space~\cite{miracle2017critical}}. Of the {thousands of HEAs investigated thus far, many exhibit properties comparable to those of conventional alloys. Some HEAs, however, do exhibit properties that are superior to their simpler counterparts}, including high strength {and} high ductility~\cite{senkov2014effect,li2016metastable}, improved fatigue resistance~\cite{hemphill2012fatigue,tang2015fatigue}, high fracture toughness~\cite{seifi2015fracture}, and high thermal stability. 
{Moreover,} the compositional (and microstructural) complexity {of} HEAs can be leveraged to {design} alloys {capable of defeating performance trade-offs currently limiting many conventional alloys}.

In fcc alloys, for example,
{low SFE tends to increase the tendency for an alloy to undergo deformation twinning, increasing the dislocation storage capacity, strain hardening rate, and plasticity -- i.e., TWinning Induced Plasticity(TWIP)~\cite{bahmanpour2011effect,wang2017hierarchical}.} Further lowering of {the SFE can promote the fcc-hcp phase transition}, resulting in {enhanced plasticity} through the Transformation Induced Plasticity (TRIP) effect~\cite{liu2019transformation}. {While SFE is not the only intrinsic property controlling 
plasticity mechanisms in fcc crystals  \cite{huang2018twinning}, its value is highly correlated with the onset of TRIP, TWIP or slip-dominated plasticity\cite{zhao2006tailoring}.} 


{The concept of tuning SFE to control the plastic deformation mechanisms in fcc HEAs has been investigated before}. 
Lu \etal~\cite{lu2018bidirectional}, for example, investigated an Fe$_{50}$Mn$_{30}$Co$_{10}$Cr$_{10}$ alloy and showed that, upon deformation, {the alloy underwent so-called a bidirectional TRIP effect. They attributed this behavior} to the unusually low value of the SFE of $\sim$6.5$\frac{mJ}{m^2}$.
 Zaddach \etal~\cite{zaddach2013mechanical} synthesized a series of alloys belonging to the CoCrFeMnNi system and determined their SFE by combining DFT calculations and microstrain measurements (via XRD), finding that Ni was positively correlated with a high SFE. Liu \etal~\cite{liu2019transformation} investigated the Fe$_{20}$Co$_x$Ni$_{40-x}$Cr$_{20}$Mn$_{20}$ system and found that increasing Co led to a decrease in SFE, which increased the tendency of the alloys to undergo TWIP and TRIP.
They also found that the drop in SFE with the addition of Co was highly correlated with the (predicted) decrease in the Gibbs free energy difference between fcc and hcp, $\Delta G^{fcc \to hcp}$. This is consistent with recent work~\cite{wei2019novel} that shows that Co-rich variants of CoCrFeNiMn underwent TWIP and TRIP-assisted deformation as a result of their lower SFE. Liu \etal~\cite{liu2018stacking} inferred the SFE of several alloys within the CoCrFeMnNi system using the weak-beam dark-field (WBDF) technique~\cite{cockayne1971measurement} and found a significance dependence of SFE on the Ni content, with lower values of SFEs promoting an increased density of deformation twins, resulting in better mechanical properties.

{\par}Computationally, Huang \etal~\cite{huang2015temperature} investigated the SFE of CoCrFeMnNi HEA using DFT calculations, de-convoluting chemical, magnetic, and strain effects. They predict a SFE of $\sim$20$\frac{mJ}{m^2}$. {The most significant contributor to the SFE was determined to be the chemical composition}.
{Zhao \etal~\cite{zhao2017stacking} investigated the SFE and the Generalized Stacking Fault Energy Surface (GSFES) for alloys in the CoCrFeMnNi and CoCrFeNiPd systems. They predicted the SFE of these alloys in two ways. The SFE was estimated directly from the GSFES, which was obtained by shearing a special quasirandom structure (SQS) approximation to a given alloy chemistry. They also predicted the SFE by using the Axial Next-Nearest-Neighbor Interaction (ANNNI) model~\cite{chaudhary2017data}, which connects the SFE to the lattice energies of a given alloy in the fcc, hcp and dhcp structures. Overall, they found good agreement between the two approaches to estimating SFE.} 

Kivy \etal~\cite{kivy2017generalized} calculated the GSFES in CoCrFeNi alloys as well as the influence of micro-alloying elements via DFT calculations. Using the theory of Tadmor and Bernstein~\cite{tadmor2004first} they further investigated the twinnability of those alloys and found that Mn, Cu and Al increased the tendency for dislocation-mediated slip and martensitic transformation, while Ti and Mo promoted dislocation glide and mechanical twinning. Ding \etal~\cite{ding2018tunable} demonstrated theoretically that chemistry alone does not determine SFE, as the latter  potentially can be tuned by controlling the short-range order (SRO). Similar to Liu \etal~\cite{liu2019transformation}, they found that SFE was highly correlated to the competition for stability between fcc and hcp configurations. Similarly, Ikeda \etal~\cite{ikeda2018impact} found that \emph{local fluctuations} in the chemistry surrounding intrinsic stacking faults resulted in significant variance in SFE. 

Zhang \etal~\cite{zhang2017origin} investigated the SFE of CoCrNi and CoCrFeNi alloys using DFT calculations of special quasi-random structures (SQS) with intrinsic faults. In their work, they report that both alloys have negative SFE, indicating that hcp configurations are more stable than fcc ones. Notably, they found significant variance in their results---SFE varied from -77 to -18 $\frac{mJ}{m^2}$ in CrNiCo---depending on the \emph{local} composition of the alloy around the intrinsic stacking fault (ISF). \emph{This points to the significant challenges of using small cell approximates to random solutions to compute highly local properties}. We note, however, that the negative SFE of CoCrNi calculated by Zhang \etal is in conflict with experimental measurements~\cite{laplanche2017reasons} that report a low, but positive SFE of $\sim$20 $\frac{mJ}{m^2}$. Ding \etal~\cite{ding2018tunable} have sought to resolve this discrepancy by invoking the effect of short range order (SRO). Indeed, they {predicted a strong influence of SRO} on the SFE in the CoCrNi system. Without invoking SRO, Zhao \etal~\cite{zhao2017stacking} showed how the SFE of CoCrNi {went from negative to positive at higher temperatures as a result of the switch (due to entropic contributions) in the relative stability of hcp- and fcc-like atomic configurations.}   {Overall, experiments and predictions tend to diverge whenever SFE is predicted to be negative.} A possible resolution to {this tension between experiments and simulations} has been put forward by Suna \etal~\cite{suna2020can}.   {They argue that the experimental inference of SFE are ultimately tied to models that are themselves \emph{biased} towards positive SFEs. This bias renders these experimental techniques unable to interpret alloys with underlying negative SFEs.}

{Experimental determination of SFE is extremely challenging~\cite{pei2018overview} and this may explain the sparseness in the experimental literature on SFEs of HEAs. To date, the vast majority of experimental studies focus on a very small number of compositions around the 'Cantor' CoCrFeMnNi system~\cite{naeem2019stacking,ding2018tunable,woo2020stacking,woo2020comparison,zhang2017dislocation}.  Even on the computational front, however, most approaches to predicting SFE in FCC HEAs center around the effects of chemical modification to the Cantor alloy group~\cite{yu2019influence,kivy2017generalized,zhang2017origin,ikeda2018impact,zhao2019effect} without explicitly considering a wider chemical space.}

Given the sparcity of the (experimental or computational) data available, {it is not surprising that there are not many examples of ML-based approaches to the prediction of SFEs in HEAs.} An exception is the work by Arora \etal~\cite{arora2020machine}, who recently demonstrated a framework to predict the SFE of Ni-Fe, Fe-Cr and Ni-Cr binary alloys. They trained ML models against a wide range of atomic arrangements set up through classical molecular dynamics simulations. The ML models were trained against specific sets of atomic pair interactions (bonds). Such models were then used to predict the SFE in higher order (i.e. ternary) systems. The data used by Arora \etal to train their models, however, were produced {by atomistic simulations driven by classical potentials there is a possibility that such calculations would not agree with \emph{ab initio}-based predictions.} {We note, however, that} ML approaches have been used before to predict SFEs, {albeit in other alloy systems}. For example, Chaudhary \etal~\cite{chaudhary2017data} developed a classifier for the SFEs in austenitic steels. More recently, Wang and Xiong~\cite{wang2020stacking} combined CALPHAD-based models and ML to predict the SFE in austenitic steels. Clearly, it is possible to develop predictive models for this important intrinsic alloy property. However, a major challenge is to produce sufficient high-quality data to develop such models.

\begin{figure}[H]
    \centering
    \includegraphics[width=0.8\textwidth]{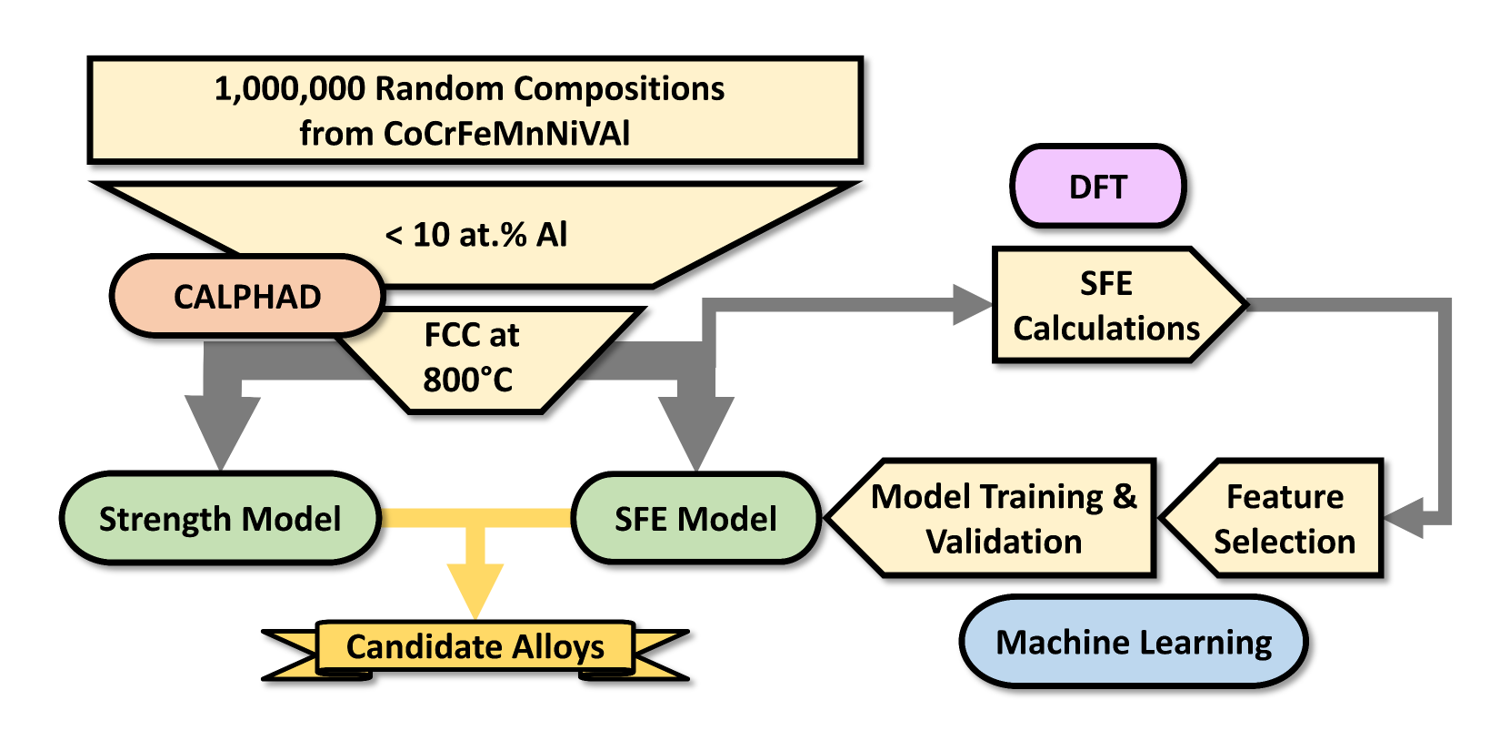}
    \caption{Alloy design framework employed in the current work. An initially vast composition set is narrowed to a few candidate alloys by applying constraints on CALPHAD-predicted properties and other properties like strength and SFE. The computational barrier of computing SFE is overcome by training Machine Learning models from a smaller set of DFT-calculated data.}
    \label{fig:framework}
\end{figure}

In this work, we put forward an approach to {\em explore} and {\em exploit} the SFE landscape in FCC HEAs. Specifically, as a major issue, we address the sparcity in the current knowledge involving the impact of chemistry on SFE   
{through a highly integrated framework that incorporates CALPHAD-based alloy analysis, DFT predictions of SFE in a finite number of alloys, and ML models capable of predicting the SFE over the entire CoCrFeMnNiV-Al FCC HEA space}. Figure~\ref{fig:framework} illustrates how this framework has been applied in the current work. An initial sampling of 1,000,000 compositions from the CoCrFeMnNiV-Al space was filtered to just those compositions that were predicted to be FCC at 800$^\circ$C.  Even this set of compositions is too large to be completely simulated with DFT.  {Therefore, a small subset of the space was selected through advanced experimental design approaches. The SFE of the selected alloys was computed using DFT methods and ML models were then generated, which could also be validated using independent DFT estimates based on SF-defected supercell with suitable configurational averages.} These models were then combined with a recently developed model for the intrinsic strength in compositionally complex FCC alloys~\cite{oh2019engineering} to explore a much larger alloy space, uncovering a small subset of alloys that are predicted to overcome the strength-ductility trade-off through the exploitation of additional plasticity mechanisms {that result from low SFEs.} {These alloys are located on the {\em Pareto front} of the strength-(target SFE) multi-objective space}. 

\section{CALPHAD Sampling}
\label{sec:calphad}


Before training a SFE model, sample compositions were drawn from the CoCrFeMnNiVAl space, which contains the Cantor system from which many common fcc HEAs originate, as well as  V and Al, which provide potential benefits in high-temperature applications and precipitate strengthening, respectively. The sampling process was initiated by generating 1,000,000 uniform random samples from the CoCrFeMnNiVAl composition space. From these sampled compositions, only those with less than 10 at.\% aluminum were kept, leaving 467,228 samples remaining. CALculation of PHase Diagrams (CALPHAD) modeling~\cite{saunders1998calphad} was then used to predict the phase stability in each composition, {using Thermo-Calc's TCHEA3 \cite{mao_tchea1_2017, chen_database_2018} database.} 
Of the 467,228 compositions tested, {\em only 36,294 -- {\bf less than 10\%} -- were predicted to be single-phase fcc at 800$^\circ$C}. {The temperature of 800$^\circ$C was selected as the target for analysis as we assumed that if an alloy is single-phase fcc at this temperature, it will likely remain so until melting. Temperatures above 800$^\circ$C could be used to solution-anneal such alloys. Thus, we assumed that the population of alloys that met this phase stability constraint could be amenable for further optimization by tuning their SFE and \emph{intrinsic} strength \emph{via chemistry.}}

{From the 36,294 remaining alloys}, an additional subset was selected to identify alloys that are suitable for high-temperature applications and additive manufacturing, as a potential manufacturing route for fcc-HEA components\cite{sun2021reducing,johnson2019assessing}. First, equilibrium solidification simulations were performed for each of the fcc samples to determine which compositions have a solidus temperature greater than 1600 K (20,541 samples) and a solidification range smaller than 100 K (35,680 samples). {Smaller solidification ranges have been shown to reduce the risk of hot cracking during additive manufacturing by minimizing the opportunity for liquid to penetrate solidified dendrites \cite{bollinghaus_hot_2008, tang_alloys-by-design_2021}.} The 20,147 compositions that met both criteria were further investigated with more time-consuming Scheil solidification simulations \cite{scheil_bemerkungen_1942}{---the underlying assumption of Scheil analysis is the lack of diffusion in the solid phase, making it compatible with far-from-equilibrium solidification processes, such as those prevalent during additive manufacturing \cite{keller_application_2017}}. Of the alloys tested, 6,935 samples had a solidification range below 100 K, while 5,590 samples had a solidus temperature above 1600 K, while 5,379 samples satisfied both criteria.

Of the 5,379 compositions determined to be suitable for high-temperature applications -- {\bf less than 1.5 \% of the total HEA space explored}, 398 were chosen for further analysis. {Their SFE was calculated using DFT calculations, as described in \cref{sec:dft}.}   {While the initial 398 compositions represent alloys suitable for high-temperature applications, an additional 100 alloys were selected so the SFE model could be trained for \emph{all} fcc alloys in the system. These additional alloys were selected by performing a \textit{k-medoids clustering} ($k=100$) on the $36,294-5,379=30,915$ fcc samples that did \textit{not} meet the solidification conditions.}   K-medoids clustering is similar to k-means in that it distributes cluster centroids across the space to minimize intra-cluster variance, but different in that cluster centroids, or medoids, must be data points themselves. As such, k-medoids clustering {can be used to generate representative subsets of a larger dataset. We note that k-medoids clustering is used as a space-filling technique, with each medoid being representative of a subset of the chemical space. Selecting alloys at random, on the other hand, would not guarantee a representative sampling of the space, particularly under sparse sampling conditions \cite{pronzato2012design,crombecq2009space}.} SFEs were also calculated for these samples to obtain a broader set of data throughout all fcc alloys in the composition space. In total, SFE values were calculated for 498 compositions in the CoCrFeMnNiVAl space. {Using this re-sampling scheme it was possible to obtain a more representative sample of the fcc HEA space, while still focusing our model over what we considered was the feasible performance region.}

\begin{figure}[htp]
    \centering
    \includegraphics[width=1\textwidth]{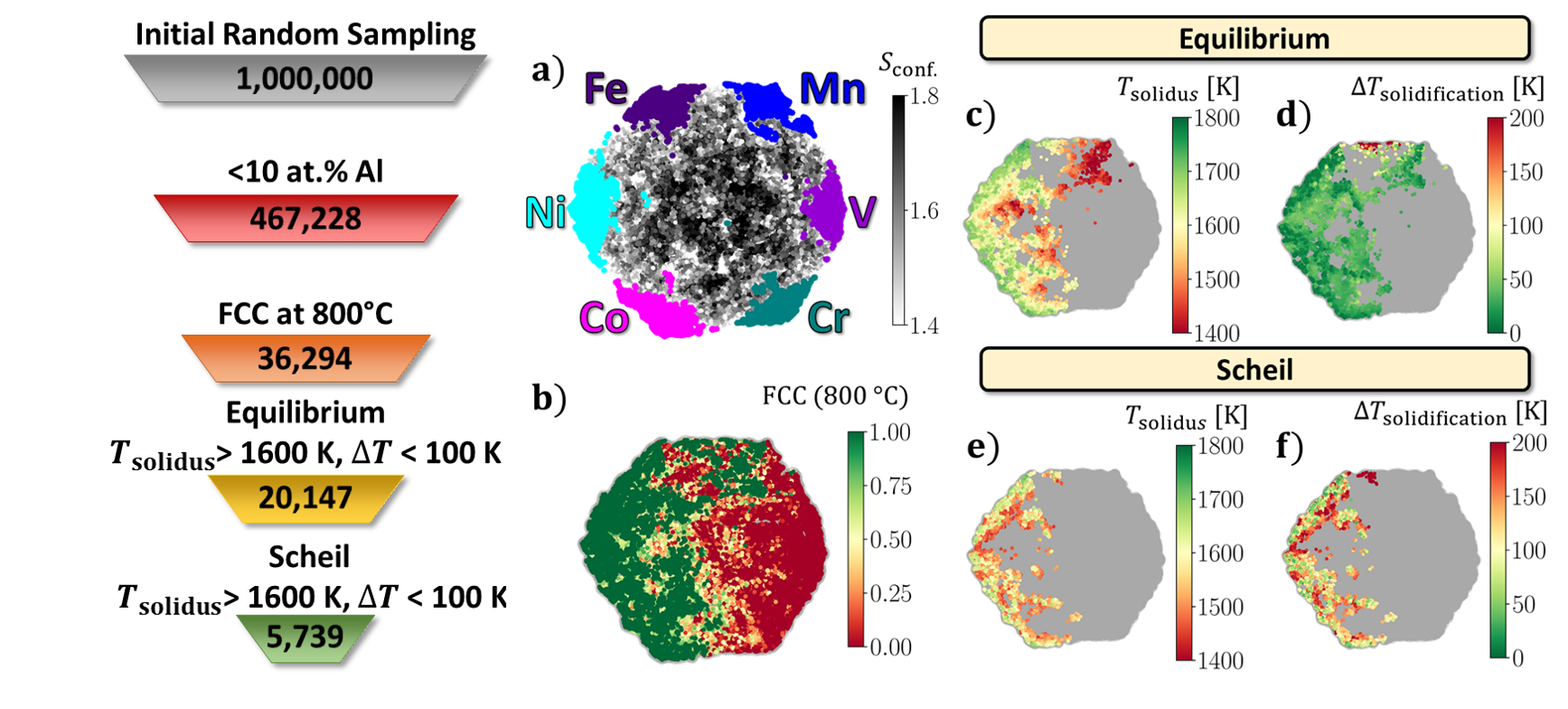}
    \caption{CALPHAD-predicted properties visualized across the CoCrFeMnNiV-Al composition space in two dimensions with t-distributed stochastic neighbor embedding (t-SNE): a) the initial CALPHAD sampling of the restricted composition space ($<$10 at.\% Al) where compositions are color-coordinated based on majority element ($>$50 at.\%) and color represents configuration entropy ($S_\text{conf.}$) for compositions without a majority element; b) the calculated equilibrium phase fraction of fcc at 800$^\circ$C; equilibrium predicted c) solidus temperature and d) solidification range; and Scheil simulated e) solidus temperature and f) solidification range. The numbers to the left of the figure represent the number of alloys that satisfied each successive constraint.}
    \label{fig:calphad_figures}
\end{figure}

Figure~\ref{fig:calphad_figures} visualizes the CALPHAD-predicted properties in the CoCrFeMnNiV-Al system. The plots shown in Fig.~\ref{fig:calphad_figures} were created via t-distributed stochastic neighbor embedding (t-SNE), a dimensionality reduction technique that can embed the 7-dimensional composition space into just two dimensions. In short, data points that are close to one another in the high dimensional space should be close in the reduced space, but the conclusions drawn from the embedding should be treated with caution due to the stochastic nature of the process. Figure~\ref{fig:calphad_figures}a shows the t-SNE embedding of the entire composition space. Note that compositions with majority elements ($>$50 at.\%) are grouped with like compositions and separated from other majority elements, resulting in a near-hexagonal distribution (there are only six possible majority elements as Al content was kept below 10 at.\%). In general, the (maximum possible) configurational entropy of compositions near the center of this hexagon is higher than the entropy of compositions near the edges.

The equilibrium mole fraction of fcc at 800$^\circ$C, as predicted {using computational thermodynamics (e.g. CALPHAD methods)} for 467,228 compositions, can be seen in Fig.~\ref{fig:calphad_figures}b. {Unsurprisingly}, trends show a strong tendency to form fcc in compositions rich in Ni and Co, as well as, to a lesser degree, Mn and Fe. In contrast, fcc is less likely to form in V- and Cr-rich  portions of the composition space. These tendencies are in agreement with the equilibrium crystal structures of the pure elements. Equilibrium solidification properties were calculated for the 36,294 compositions predicted to be single-phase fcc and can be seen in Figs.~\ref{fig:calphad_figures}c and d. The solidus temperatures of most compositions are consistently above 1600 K, except for Mn-rich compositions which are about 200 K lower. Similarly, solidification ranges are mostly below 100 K, except for some compositions with high Mn contents. Both of these observations can be explained by Mn's lower melting temperature (1246 K) in comparison to Fe, Ni, and Co (1583 K, 1455 K, and 1495 K respectively). The Scheil solidification behavior was calculated for the 20,147 compositions with equilibrium solidus temperatures greater than 1600 K and solidification ranges lower than 100 K and are visualized in Figs.~\ref{fig:calphad_figures}e and f. There is a general trend for the solidus temperature to be lower and the solidification range to be higher as compositions move away from Fe, Ni, and Co. This can be explained by the similarity of the melting temperatures of those elements and the difference in melting temperatures introduced by the other potential alloying elements: Mn and Al with lower melting temperatures and V and Co with higher. 

Even considering the uncertainties and likely inaccuracies of the thermodynamic database used, it is evident from the analysis above that the \emph{feasible} HEA space, at least in the case of the fcc "Cantor+VAl" region considered here, is very small relative to the entire {HEA space}. This result runs counter to the conventional implicit assumption as to the vastness of the HEA space but is in line with what is known from other metal alloy systems: feasible alloys tend to be located in very narrow composition ranges. Given the considerable challenge in exploring the HEA space, though, such "filtering" of the feasible space is necessary.

\section{Density-Functional Theory Calculations: Stacking Fault Energy}
\label{sec:dft}
Having selected a group of alloys that belonged to the feasible region in the Cantor+V-Al space, their stacking fault energy (SFE) was calculated as described here. {A stacking fault} can be simply understood as the planar defect inserted in the {-A-B-C-A-B-C-A-} stacking sequence in a face-centered cubic (fcc) crystal to {-A-B-C|A-B-A-B|C-A-}. Then, the SFE corresponds to the energy difference between a crystal with a stacking fault and the perfect fcc stacking sequence. There are a number of approaches based on DFT methods that can be used to calculate the SFE of crystals~\cite{pei2018overview}. One of the most common approaches is to simply shear a crystal along the (111) plane in order to calculate the Generalized Stacking Fault Energy Surface (GSFE), which corresponds to the 'energy landscape' seen by the crystal as it is being sheared. While implementation of such calculation is relatively straightforward, the calculation of SFE for alloys with arbitrary compositions is very challenging. Alloy disorder in finite crystals (under periodic boundary conditions) can be simulated using special quasi-random structures (SQSs) \cite{van2013efficient} or equivalent rapid optimization using Super-Cell Random APproximates (SCRAPs) \cite{singh2020}. A major challenge associated with the use of SQS-type methods is that it is {very difficult} to design periodic supercells with arbitrary {alloy compositions}. {Even when done, it comes at the cost of generating large supercells that are extremely costly to simulate within DFT}. Even when supercells of the target chemistry are designed, however, explicit shearing of small SQSs have already been shown to result in significant variance in the computed SFE due to the very strong dependence of the energetics of a given crystal on the \emph{local} chemical composition along the stacking fault plane~\cite{ikeda2018impact}. 

\begin{figure}[htp]
    \centering
    \includegraphics[width=0.5\textwidth]{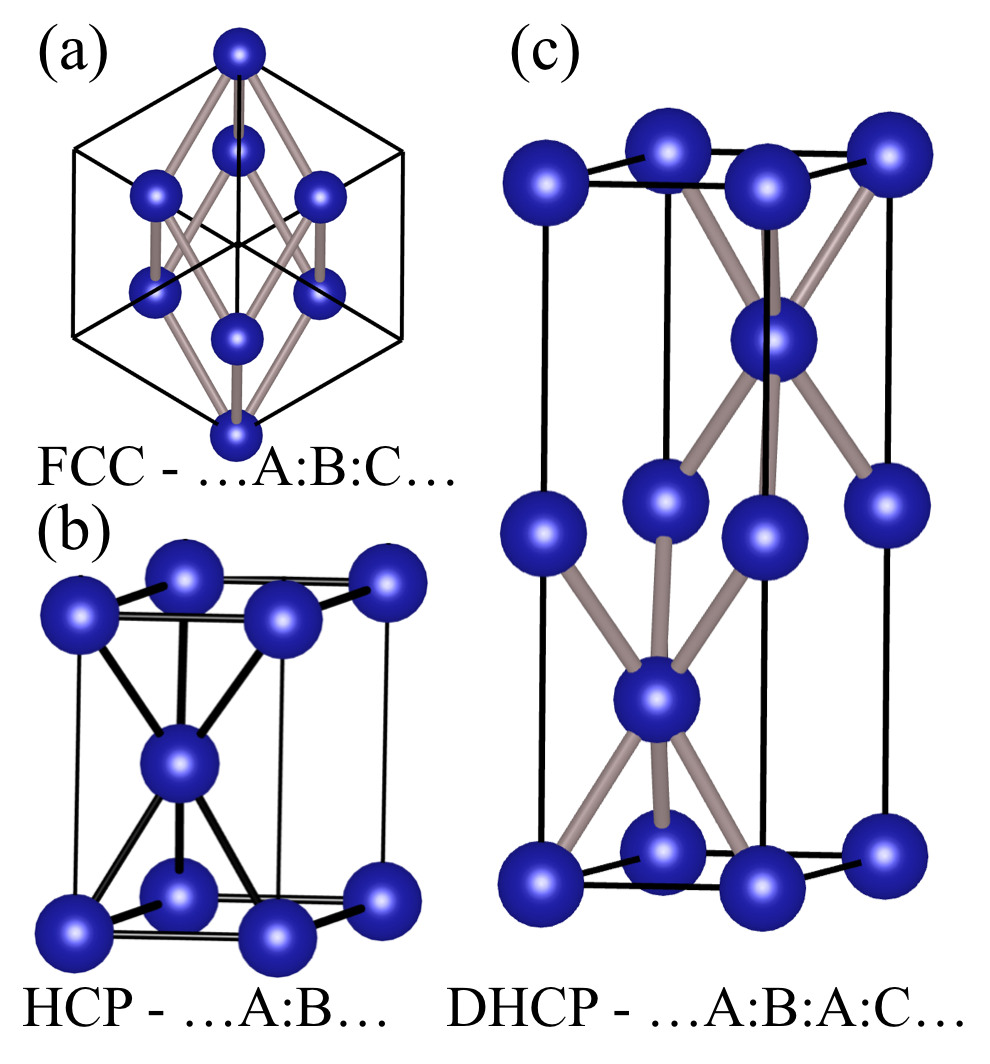}
    \caption{a) Primitive cells of fcc (...A:B:C...), hcp (...A:B...), and dhcp(...A:B:A:C...) unit cells used for total energy calculations to estimate stacking-fault energy in the ANNNI model.} 
    \label{fig:annni_model}
\end{figure}

To avoid this roadblock, we used instead the axial next-nearest-neighbor Ising (ANNNI) model \cite{denteneer1991energetics,denteneer1987stacking} to calculate {the SFE of selected alloys}, as in Fig.~\ref{fig:annni_model}. In the ANNNI model with  pairwise (Ising) interactions, the stacking sequence can be modeled as a one-dimensional crystal, whose energetics are governed by pair interactions of extended ranges. After mapping the stacking sequence along directions normal to the (111) plane in face-centered-cubic (fcc; $...A:B:C:A...$), hexagonal closed-packed (hcp; $...A:B:A:B...$), and double hexagonal closed-packed (dhcp; $...A:B:A:C...$) crystals, the (intrinsic) SFE can be computed using the analytic energy-difference expression:
\begin{equation}\label{ANNI-SFE}
\gamma_\text{ISF}^\text{ANNI} = \big(E_\text{hcp}+2E_\text{dhcp}-3E_\text{fcc}\big)/A  ,
\end{equation}
\noindent where $A$ is the stacking fault area along the (111) fcc plane. Calculated from DFT, $E_{hcp}$, $E_{dhcp}$, and $E_{fcc}$ {correspond to} the total energy-per-atom in each crystal structure with a configurational average over {atomic configurations}  The total energy in ANNNI model SFE expression (eq. 1) was calculated directly using the DFT-based electronic-structure Green's function method (DFT-KKR-CPA) for disordered alloys ~\cite{johnson2008mecca,johnson1986density,johnson1993inclusion,alam2009optimal,alam2012structural,johnson1985self}  
{that averages over the infinite set of environments within a single-site approximation for a given structure (e.g., fcc, hcp, or dhcp) during the DFT charge self-consistency. We used spin-polarized setting in all our calculations due to presence of magnetic elements (Fe, Co, Cr, Ni, Mn) in HEAs.} (This is in contrast to an SQS that has a single configuration in a finite cell).  { In addition, providing an independent validation, the DFT-KKR-CPA can also address any defected crystal structure.  So, similar to that for elemental metals,  a direct SFE estimate is 
\begin{equation}\label{KKRCPA-SFE}
\gamma_\text{ISF}^\text{direct} = \big(E_\text{SF} - E_\text{fcc}\big)/A  ,
\end{equation}
where configurational averaging is performed for 1-atom fcc and an 11-atom intrinsic SF cell (i.e., 11 (111)-layers), rather than 2-atom hcp and 4-atom dhcp for an ANNI model estimate. }

{The KKR-CPA Green's function technique  is a fully variational, exact DFT method, unlike historical tight-binding versions, and it  can be extended beyond single-site average, if needed,  using a cluster generalization of the CPA \cite{PhysRevB.72.113105}. The KKR-CPA has been used to explore quantitatively numerous  complex chemically in solid-solution alloys, including quantum critical points \cite{PhysRevLett.107.206401}, vacancy-mediated phase selection \cite{singh2020vac}, and SFE variation in martensitic transformations in multi-principle-element alloys \cite{PhysRevLett.127.115704}. 
In brief, the KKR-CPA permits the structural defect to be added to the crystal structure and then concomitantly with the DFT charge self-consistency an average is performed simultaneously over all configurations -- not just for one selected configuration as in SQS supercell methods.  When modeling alloy disorder, the KKR-CPA approach is better than the DFT+SQS technique because the former can account for arbitrary chemical compositions in a self-consistent manner, whereas the composition resolution in SQS-based random alloy is limited by the size of the supercell used.}

The Perdew-Burke-Ernzerhof (PBE) exchange-correlation functional was included through \texttt{libXC} libraries~\cite{marques2012libxc,perdew1996generalized}. The Monkhorst-Pack k-point method was used for the Brillouin zone (BZ) integration~\cite{monkhorst1976special}. K-meshes of $18\times18\times18$, $18\times18\times6$, and  $18\times18\times6$ dimensions were used to calculate the properties of the fcc, hcp, and dhcp phases. The Birch-Murnaghan equation of state was employed to fit the energy-volume of fcc, hcp, and dhcp phases to determine their equilibrium lattice constants~\cite{birch1947finite}. The ideal c/a ratio in the hcp phase was used in all calculations---this simplification can be relaxed by minimizing the crystal energies of hcp cells with respect to c/a ratio, {but this would significantly increase the cost of the computational workflow with only a minimum gain in accuracy. Such an accuracy gain would be eclipsed by the uncertaity in the resulting ML models}

{\it Validation of SFE}:~To validate DFT-predicted SFEs, we calculated {the SFE of some metals and alloys for which the SFE has been determined experimentally, such as Al, Ni, austenitic steel, and Fe-Mn-Co-Cr-based HEAs}. {In the case of elemental solids Ni and Al}, the DFT-predicted SFE are 130 mJ/m$^2$ and 136 mJ/m$^2$, respectively, in good agreement with experimental values of Al (120-130 mJ/m$^2$\cite{doi:10.1080/14786437708232942,DATTA2009124}) and Ni (125 mJ/m$^{2}$  \cite{1983JAM}).   The SFE  for Fe$_{40}$Mn$_{40}$Co$_{10}$Cr$_{10}$ solid-solution alloy {has been estimated as} 17$\pm${4} mJ/m$^2$  \cite{picak2019anomalous, PhysRevLett.127.115704}, {in good agreement with our predicted ANNI model (eq. 1) value of 22.2 mJ/m$^2$ and our direct SFE (eq. 2 validation) value of 22.3 mJ/m$^2$. Ausenitic steels in the Fe-Mn-Al-Si system have been investigated extensively, with a number of experimental determinations of their SFE~\cite{chaudhary2017data}. Here, we focused on the Fe-24.7Mn-2.66Al-2.92Si alloy for which Pierce et al. measured a SFE of 16$\pm$4 mJ/m$^2$~\cite{PIERCE2012753}.
Similarly for equiatomic NiCoCr, ANNNI estimated SFE  (20.7 mJ/m$^{2}$ from eq. 1) was found in good agreement with direct DFT SFE (18.5 mJ/m$^{2}$ from eq. 2) and experiments (22$\pm$5 mJ/m$^{2}$) \cite{laplanche2017reasons}. From this limited comparison, we estimate that the discrepancy between our DFT-predicted SFEs and what is measured experimentally -- noting again that only \emph{positive} SFEs can be inferred experimentally -- is within $\sim$4~$\frac{mJ}{m^2}$.}

\begin{figure}[htp]
    \centering
    \includegraphics[width=\textwidth]{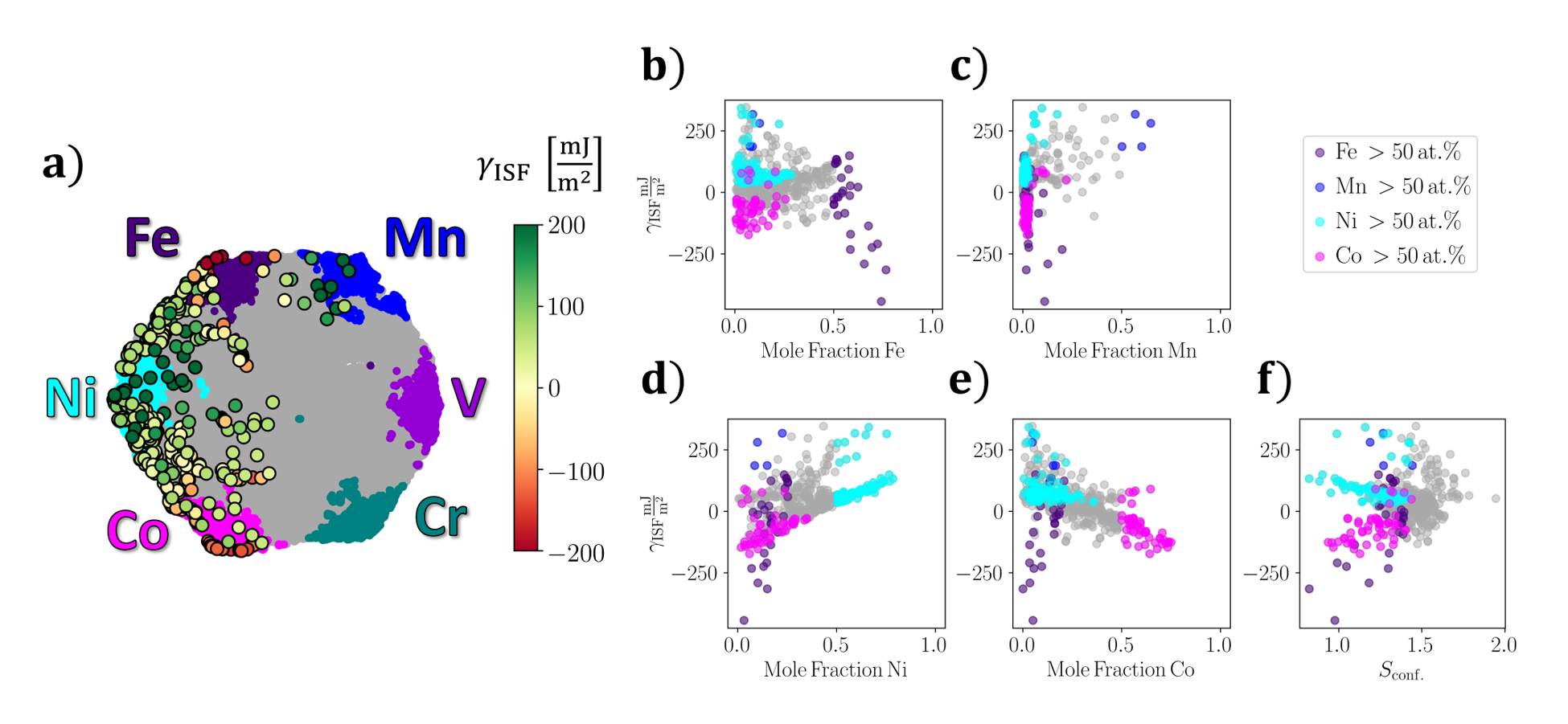}
    \caption{a) DFT-predicted stacking-fault energy visualized on a t-SNE embedding of the CoCrFeMnNiVAl composition space. b-e) SFE plotted against each of the four fcc-forming elements. f) SFE plotted against configurational entropy.}
    \label{fig:sfe_vs_composition}
\end{figure}

{\it ANNI-based SFE prediction}:~Figure~\ref{fig:sfe_vs_composition}a shows the calculated SFE of the 498 compositions chosen to represent fcc-forming alloys in the CoCrFeMnNiVAl composition space. Figures ~\ref{fig:sfe_vs_composition}b-e display the same SFE values against each of the four elements that appear as a majority in at least one of the fcc compositions. SFE exhibits a positive correlation with Mn and Ni, while the correlation is negative in the case of Fe or Co. The two strongest composition correlations are associated with Ni and Co. As discussed in the introduction, these results are supported by the work of Zaddach \etal~\cite{zaddach2013mechanical} in the case of Ni and Liu \etal~\cite{liu2019transformation} in the case of Co; both relationships can be largely explained by each element's tendency to form fcc (Ni) or hcp (Co). Figure~\ref{fig:sfe_vs_composition}f plots SFE against the (maximum possible) configurational entropy of each composition ($S_\text{conf} = -\sum_{i=1}^{n}x_i\ln{x_i}$). Unsurprisingly, no significant trend can be observed as there is no plausible connection between compositional complexity (indicated by this maximum configurational entropy) and SFE.    Note that the tendency of SFE to be less extreme (not too positive or too negative) as configurational entropy increases is simply the result of the \emph{cancellation} of {effects as the negative or positive contributions by Ni, Co or other constituents interfere with each other.}

\subsection*{Electronic origin of low and high SFE}
\begin{figure}[H]
    \centering
    \includegraphics[width=\textwidth]{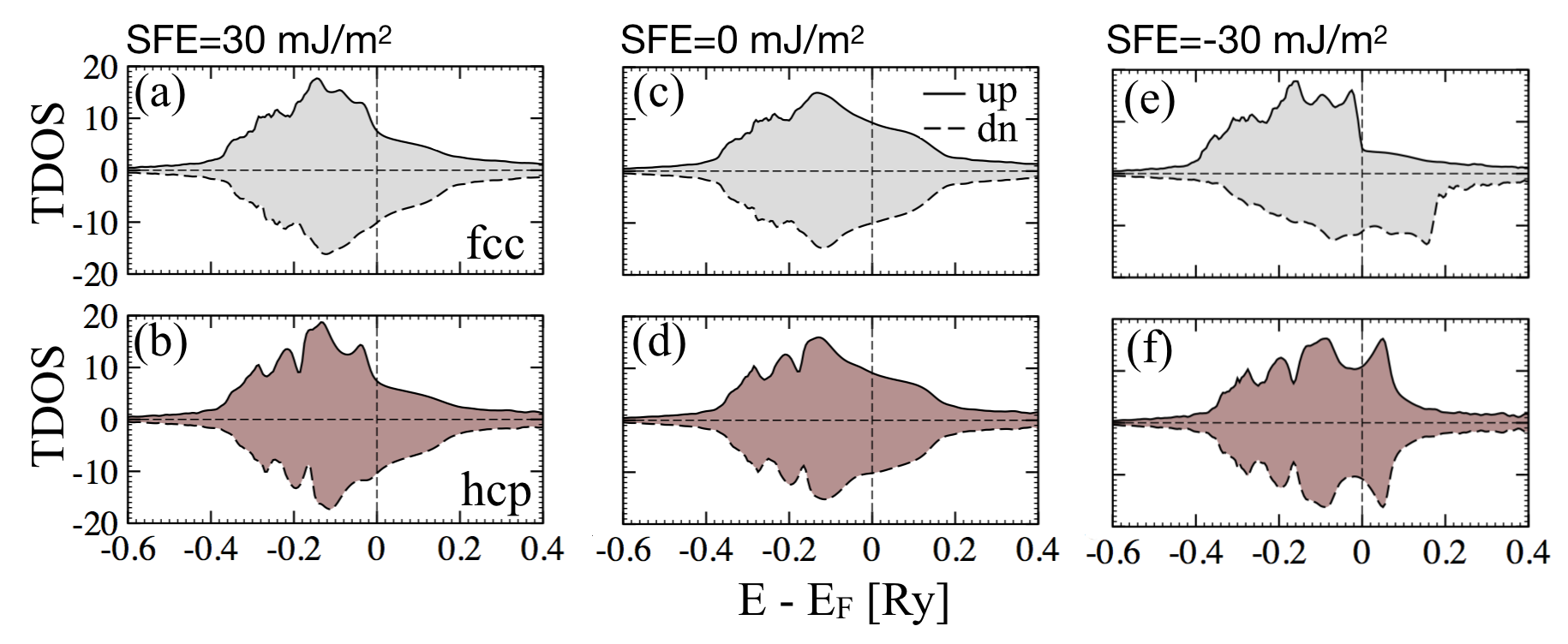}
    \caption{The total density of states (DOS) of selected fcc (top-panel) and hcp (bottom-panel) compositions with (a,b) high (30 mJ/m$^2$; Fe$_{3.188}$Mn$_{9.599}$Ni$_{66.272}$Co$_{3.035}$Cr$_{3.902}$V$_{12.674}$Al$_{1.33}$), (c,d) zero (0 mJ/m$^2$;  Fe$_{1.253}$Mn$_{0.581}$Ni$_{39.153}$Co$_{21.784}$Cr$_{31.251}$V$_{3.66}$Al$_{2.318}$), and (e,f) low (-30 mJ/m$^2$;  Fe$_{76.443}$Mn$_{1.76}$Ni$_{14.954}$Co$_{0.097}$Cr$_{2.247}$V$_{2.714}$Al$_{1.785}$) stacking-fault energy (SFE). The Fermi energy (E$_{F}$) in the total DOS is set at energy zero. The total DOS clearly reflects the signature of low/zero/high SFE through change in structure of electronic-states at E$_{F}$.}
    \label{fig:tdos_dft}
\end{figure}

{As mentioned above,} the stacking-fault energy (SFE) is a crucial quantity that links macroscopic mechanical properties such as twinning induced plasticity (TWIP) or transformation induced plasticity (TRIP) to atomic configurations in a crystal. Therefore, it would be interesting to understand the electronic-structure origin of high or low SFEs in randomly disordered HEAs. We plot {the} total density of states (DOS) for fcc/hcp HEA compositions in Fig.~\ref{fig:tdos_dft} selected from three extreme SFE zones, i.e., high (+30 mJ/m$^{2}$), zero (0 mJ/m$^{2}$), and low (-30 mJ/m$^{2}$). The {DFT-KKR-CPA results provide an {\it ab initio} description of disorder effects on the underlying configurationally-averaged electronic structure  that directly} relates to change in planar fault energy, i.e., SFE. Both structure and number of electronic states at the Fermi energy E$_{F}$ have been discussed in previous work \cite{PhysRevMaterials.2.055004} relating this to structural as well energy stability.  When talking about electronic-structure, {the presence of a pseudo gap} (a valley near E$_{F}$ in density of states) {is an indication for structural as well as chemical stability} of an alloy~\cite{PhysRevMaterials.2.055004}. {This stability can in turn originate from multiple electronic mechanisms, such as resonance among $d$-states of alloying elements, charge transfer, and/or electronic state  hybridization~\cite{PhysRevB.27.2005, PASTUREL1985177}}.

Hume-Rothery, based on empirical reasons, suggested that alloys with smaller conduction band width should have higher stability, {as a narrower width suggests higher filling, therefore, increased stability\cite{hume1963electrons}}. However, this should be seen together with idea of a pseudo-gap in the DOS, since even for filled conduction bands, metastability can arise if the DOS has a peak at E$_{F}$~\cite{PhysRevMaterials.2.055004}. While scanning total DOS with high (+30 mJ/m$^{2}$) to zero (0 mJ/m$^{2}$) to low (-30 mJ/m$^{2}$) SFE, we also observe that {the width of the conduction band of fcc phase} in Fig.~\ref{fig:tdos_dft} slowly increases for positive-to-zero SFE and it doubles for negative (low) SFE. For example, the total DOS in Fig.~\ref{fig:tdos_dft}a-b {in the case of the high SFE (+30 mJ/m$^{2}$) alloy} shows {an increased disorder-induced broadening in fcc relative to hcp due to an increased charge sharing among the states near Fermi energy (E$_{F}$) in fcc}.  {Our charge analysis further suggests that Fe/MnCr/V/Al in the fcc phase lose electrons (e$^{-} \sim$ 0.0015) compared to hcp, whereas the Co in fcc phase gains charge ($\sim$ -0.0124 e$^{-}$) compared to hcp (Ni remains almost neutral because of its filled d-state).}

This analysis also agrees well with the stability of all three phases where the formation energy difference of the fcc phase, when compared to hcp phase, goes positive, i.e., $\Delta {E^{fcc-hcp}_{form}}$=$-1$ mRy, $0.02$ mRy, $+10$ mRy. Eberhart et al. found that DOS structure at E$_{F}$ is also related to mechanical behavior, e.g., higher self-diffusion for higher DOS at E$_{F}$ was found to lower creep and suggested to enhance ductility of disorder alloys \cite{Eberhart1993}. Furthermore, Collings et al. showed that if E$_{F}$ collides with a pseudo-gap, it increases stiffness in lattice (due to $d$-state band-filling) and this improves the elastic moduli~\cite{COLLINGS1973437}. For multi-principal element alloys, we show that the electronic structure (density of states) also relates with {the} planar-defect energy of fcc alloys.

\section{Machine Learning models for SFE}
\label{sec:ML}

Even though the ANNNI-based approach in \cref{sec:dft} used to calculate the SFE is less expensive than an explicit calculation of the GSFE, calculating this quantity for the $\sim$5,000 alloys deemed feasible in \cref{sec:calphad} would have been computationally costly. {Moreover, relying on explicit calculation of SFE (using any DFT-based approach) would have precluded us from exploring wide regions of the chemical space.} To facilitate the use of SFE as an alloy design parameter, we thus proceeded to generate machine-learning (ML) models capable of connecting alloy chemistry to SFEs.

\subsection{Featurization}
\label{sub:featurization}

Two distinct feature sets could be used to train models of SFE: one based purely on the elemental composition of the alloys and another selected from material properties estimated from those of the pure elements. Using composition as a feature set is a natural analogy to the DFT calculations, as composition was the only varied input when calculating SFE. Composition is also likely to be used to query surrogate models in an alloy design scheme, as new compositions are tested for targeted SFEs. However, using material properties as features has several advantages over using composition directly. Firstly, SFE might be more correlated with {certain intrinsic properties} than with composition, meaning that features based on properties might lead to more accurate models. Abstracting features to elemental properties also enables the model to \emph{potentially} predict SFEs for compositions outside the original CoCrFeMnNiVAl composition space. While extrapolating to compositions far outside the training space might be inaccurate, accurate predictions might still be possible for alloys that contain only small amounts of elements that were not included in the training data. Lastly, models based on material properties could provide some insight on the underlying relationships between basic \emph{intrinsic} material properties and Stacking Fault Energy. For these reasons, property-based features were selected to train the surrogate models.

The process of selecting material properties as model features began with determining 17 elemental material properties that can be easily estimated for arbitrary compositions as a function of pure element properties and are likely correlated with SFE. These elemental properties include various simple atomic, electronic, structural, and elastic properties as well as properties more specifically tied to Stacking Fault Energy. Basic atomic properties include the atomic weight, the density, the atomic planar density, metallic radius, melting point, specific heat, and the total number of electrons. Other electronic properties like the number of valence electrons, ionization energy, Pauling and Allen electronegativies were also considered as well as structural properties like the shear modulus and elastic constants ($C_{11}$,$C_{22}$,$C'$). Lastly, the total energy difference (i.e. lattice stability) between the HCP and FCC structures ($E^\text{hcp-fcc}$) was considered, given the importance of this phase transition to stacking fault formation. Pure element values of $E^\text{hcp-fcc}$ were obtained from Wang \etal~\cite{wang_ab_2004}, who employed DFT (using VASP with PAW-GGA potential), as well as the Scientific Group Thermodata Europe (SGTE)\cite{dinsdale_sgte_1991} which used CALPHAD methods to estimate the lattice stability between hcp and fcc phases.

The initial set of 17 pure element properties were mapped into features by computing the compositionally-weighted mean (avg) and variance (var) for each property:
\begin{equation}
    P_\text{avg} = \frac{1}{\sum^n_{i=1}x_i} \sum^n_{i=1}x_iP_i
\end{equation}
\begin{equation}
    P_\text{var} = \frac{1}{\sum^n_{i=1}x_i} \sum^n_{i=1}x_i(P_i-P_\text{avg})^2  ,
\end{equation}
where $x_i$ is the composition of element $i$ in an $n$-component alloy and $P_i$ is the value of the arbitrary property $P$ for the pure element $i$. Subsets of the resulting 34 features were used to train each of the surrogate models, as described below.

\subsection{Machine Learning: GPR and SVR}
\label{sub:ml_gpr_svr}
\subsubsection*{Model Descriptions}

Surrogate models of the DFT-calculated SFEs were created using a variety of regression modelling techniques. Each technique employs machine learning so that surrogate models would be fast to evaluate, but also reasonably accurate over a broad 7-element composition space with only about 500 training points. 

The first technique employed is \textit{Gaussian Process Regression} (GPR). One advantage of this technique is that it generally interpolates between, and is fit exactly to, the training points. This behavior is desired in surrogate modeling as the high fidelity model---DFT in this case---can be treated as ground truth and trusted to not possess aleatoric uncertainty. Furthermore, Gaussian process models produce an uncertainty estimate that scales with distance from the training points---i.e. the largest uncertainty in the predictions corresponds to regions in the design space furthest from the sampled points---, \emph{enabling uncertainty quantification in future design endeavors that employ the surrogate model}. Such predicted uncertainty can also be used to construct experiment utility functions in closed-loop Bayesian Optimization (BO) schemes for materials discovery and optimization\cite{talapatra2018autonomous,talapatra2019experiment,arroyave2019systems,solomou2018multi}. {Three kernels were tried and compared to arrive at a best model: Radial Basis Function (RBF), Mat\'{e}rn, and Rational Quadratic. The Matern kernel is a generalization of RBF with an additional parameter to vary function smoothness. The Rational Quadratic kernel is another derivative of RBF and combines an infinite sum of RBF kernels via a scale mixture parameter. Both of these kernels are less prone to sample bias than RBF as they are less smooth and can adjust smoothness in oversampled areas of the space.}

In addition, \textit{Support Vector Regression} (SVR) was used to train models given its own unique benefits. While Gaussian Processes use all training data in the final model, Support Vector models use only a subset of training data--i.e. \textit{support vectors}---in the ultimate regression model. This makes SVR models more memory-efficient and less susceptible to sampling biases in the training data. However, uncertainty quantification is less straightforward with such models. {Three kernels were also tried and compared to arrive at a best SVR: Radial Basis Function (RBF), polynomial, and linear.}

\subsubsection*{Feature Selection}
 
A set of 34 property-based features were {downselected to reduce the dimensionality of the input space}. This feature selection process began by rescaling each feature 
from 0 to 1. Then, the Mutual Information ($I$)\cite{vergara2014review} was calculated between the DFT-predicted SFE and each feature. Unlike the Pearson correlation coefficient, mutual information can capture relationships between variables beyond those that are strictly linear. Features were then ranked in descending order of $I$. Features with the highest $I$ had {the strongest correlation with SFE} and were thus most likely to be informative to an SFE model.The pairwise correlation coefficients of all 34 features were also calculated to identify potential redundancies. The final feature set was obtained by first using k-medoids clustering\cite{madhulatha2011comparison} to identify 200 compositions that were evenly spread across the training set. The 34 property-based features at these compositions were then sorted by descending $I$ with SFE. Features were then chosen by descending $I$, \emph{as long as they did not have an absolute correlation coefficient higher than 0.5 with an already chosen feature}.

\begin{figure}[htp]
    \centering
    \includegraphics[width=\textwidth]{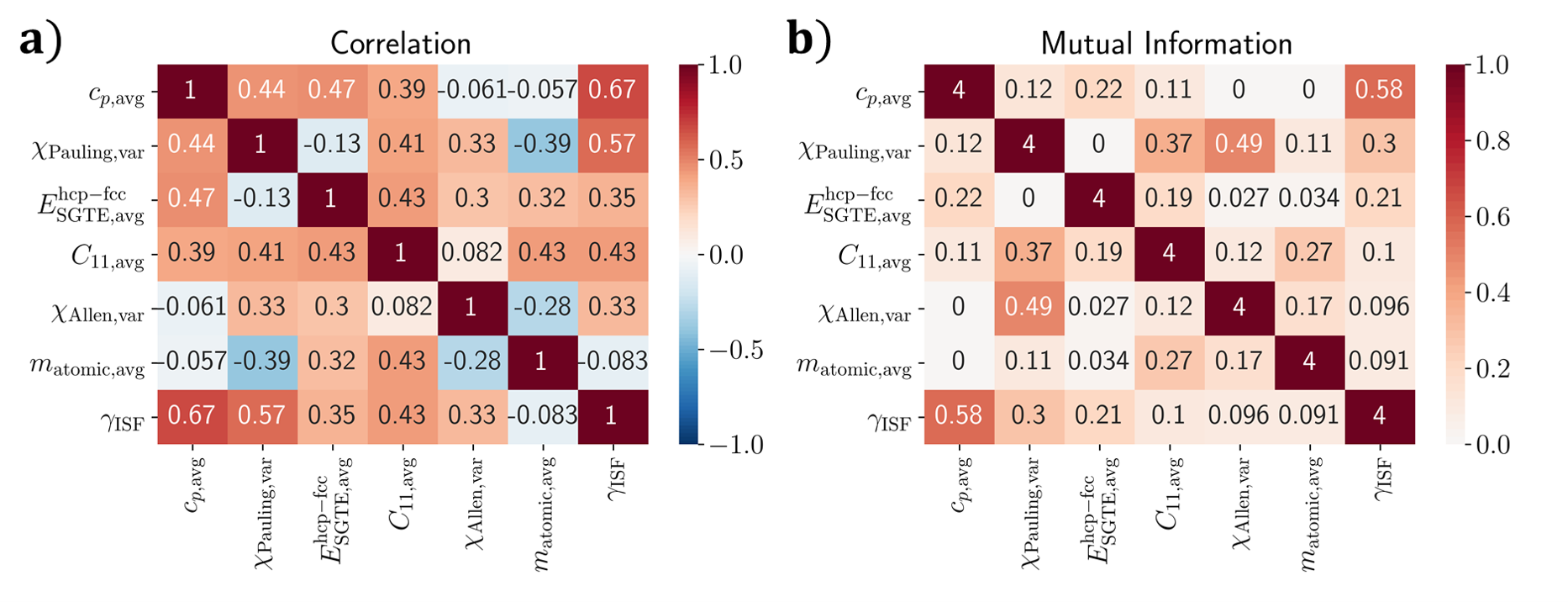}
    \caption{Matrices displaying pairwise a) Pearson correlation coefficients and b) mutual information between the six selected model features and SFE.}
    \label{fig:feature_matrices}
\end{figure}

{With this process, the chosen features were ensured to have strong correlation} with SFE, \emph{while also remaining marginally independent from each other}. In contrast, feature selection based exclusively on rank-ordering mutual information scores would have resulted in a poorly predictive model as its effective dimensionality would have been significantly reduced. The final six chosen features were the specific heat average ($c_{p,\text{avg}}$), the variance in Pauling and Allen electronegativities ($\chi_{\text{Pauling},\text{var}}$ and $\chi_{\text{Allen},\text{var}}$), the average CALPHAD-predicted lattice stability  ($E^\text{hcp-fcc}_{\text{SGTE},\text{avg}}$), the average elastic constant in the $11$ direction ($C_{11,\text{avg}}$)  and the average atomic mass ($m_\text{atomic, avg}$). Figure~\ref{fig:feature_matrices} shows the pairwise correlation and mutual information matrices with the chosen features and SFE. The average specific heat ($c_{p,\text{avg}}$) and variance in Pauling electronegativity ($\chi_{\text{Pauling},\text{var}}$) both have correlation coefficients higher than 0.5 with SFE and have the highest MI values with SFE. The selected features are also more independent than required by the selection process,{ as no feature pair within the selected subset has a correlation higher than 0.5}


\subsubsection*{Model Training and Validation}
Before training any model, SFE values were scaled by the overall minimum and maximum values of the training set so that the scaled values ranged between 0 and 1. Property-based features were also rescaled, while composition features naturally fall within the desired range.

The (lack of) diversity in the training set was of particular concern when training the machine learning surrogates due to the imbalanced spread of training points in the composition space. {The imbalance arises because} a large majority of the training points (398 compositions) came from those fcc alloys that satisfied the solidification conditions {relevant to} high temperature (HT) applications  (about 15\% of all fcc alloys). The remaining 100 compositions were taken from the roughly 85\% of fcc alloys that did not meet HT constraints. Such an imbalanced training set could negatively affect the accuracy of the models in the undersampled regions. To address the issue, k-medoids clustering was applied to the selected feature space,{\emph{over the entirety of the fcc alloy space}}, {to choose more representative subsets, irrespective of the number of alloys belonging to each class (HT vs non-HT fcc alloys)}. The number of medoids chosen (i.e., the value of $k$) was varied to determine which subset size produced the best model.  

Models were validated by computing two sets of error metrics: a \textit{test} error and a \textit{cross-validated} (CV) error. The test set error was computed by testing the trained model on all data points not used for training (i.e., those not chosen to be medoids). The cross-validated error was calculated by performing 10-fold cross-validation on the training set (i.e. the chosen medoids). Total errors were calculated on the combined test and cross-validated residuals. Hyperparameters for the Support Vector models were optimized with a single independent cross-validated Bayesian optimization. Because the k-medoids algorithm and the models themselves rely on stochastic initializations, the training process for all models was repeated ten times with different random seeds.

{For all cases, the models with the highest total $R^2$ at the value of $k$ (i.e. size of training set) with the highest average total $R^2$ were chosen to be the best models for each technique. Of the three kernels tested for GPRs, the Mat\'{e}rn kernel produced the most consistent results and the best model with a Total $R^2$ of 0.939 at a training size of 320 medoids. The Rational Quadratic Kernel produced similar results with a best Total $R^2$ of 0.927 at a training size of also 320 medoids. However, the accuracy of the RBF kernel reduced with sample size and produced a best Total $R^2$ of only 0.863 at a training size of 180 medoids. The susceptibility of the GPR trained with the RBF kernel to sample bias is likely due to its smoothness and discussed further in Appendix C. In contrast, the RBF kernel exhibited the best overall accuracy for the SVRs with a Total $R^2$ of 0.926 at a training size of 400 medoids. This was significantly better than the polynomial and linear kernels with best Total $R^2$'s of 0.874 and 0.615, respectively. The SVR with the RBF kernel was less susceptible than the GPR counterpart to sample bias because SVRs seek to keep the residuals of all training data below a specified threshold, which reduces the tendency to sacrifice residuals on the minority of data for improved performance on the majority.}

\begin{figure}[htp]
    \centering
    \includegraphics[width=\textwidth]{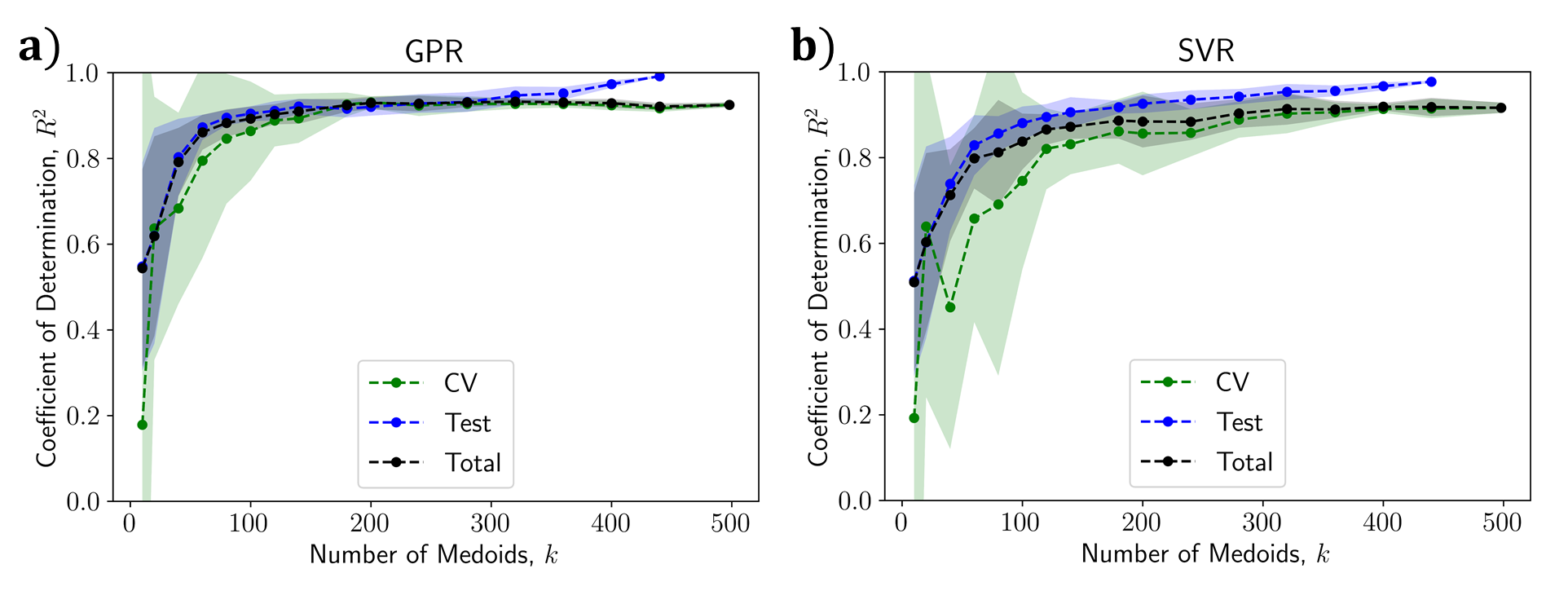}
    \caption{Coefficient of Determination ($R^2$) of each trained model versus the number of medoids selected for training ($k$) for both a) Gaussian process {(Mat\'{e}rn kernel)} and b) support vector regressors {(RBF kernel)}. Plots show the 95\% confidence intervals of ten random runs and $R^2$  is listed independently for the cross-validated training data (CV), the untrained data (Test), and all data together (Total). }
    \label{fig:models_vs_k}
\end{figure}

Figure~\ref{fig:models_vs_k} displays the coefficient of determination, $R^2$, of the trained Gaussian Process Regressors (Mat\'{e}rn kernel) and Support Vector Regressors (RBF kernel) as a function of the number of medoids or data points in the training set for all ten random runs. {Figure~\ref{fig:models_vs_k} also depicts the uncertainty due to random seeding (i.e. the random selection of training data via k-medoids) via 95\% confidence intervals. Note that the uncertainty is much higher with a smaller number of medoids. The uncertainty in the both models reduces to be very small as the number of medoids increases to the maximum.}

\begin{figure}[htp]
    \centering
    \includegraphics[width=\textwidth]{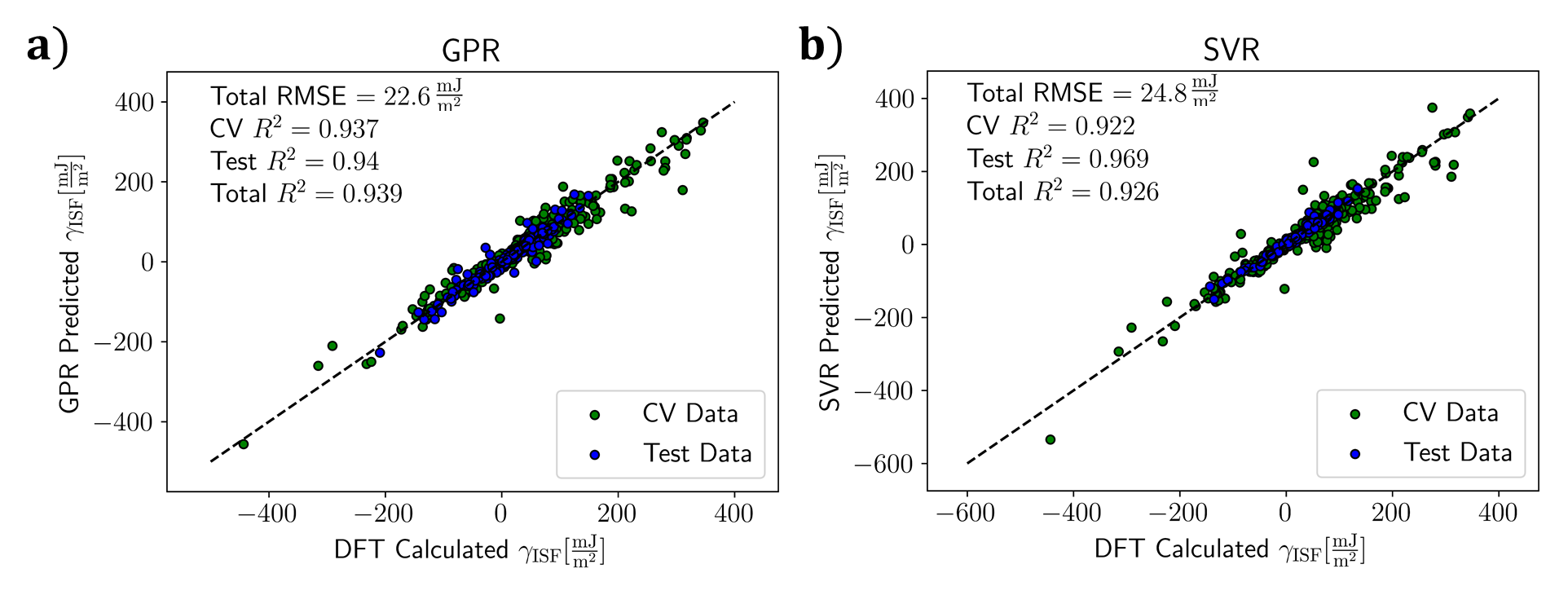}
    \caption{Predicted versus actual stacking fault energy for the best a) Gaussian process {(Mat\'{e}rn kernel)} and b) support vector regressors {(RBF kernel)}. Predicted values for the training data were obtained via cross-validation (CV Data). Models were also tested against data points that were not chosen as medoids (Test Data).}
    \label{fig:best_models}
\end{figure}

 Figure~\ref{fig:best_models} displays the predictions of SFE against their actual values for the best GPR and SVR. The best GPR {used the (Mat\'{e}rn kernel) and was trained with 320 medoids: 222 from the 398 HT compositions and 98 from the 100 non-HT fcc compositions}. Meanwhile, the best SVR {used the RBF kernel} and was trained with 400 medoids: 300 of the 398 HT compositions and all 100 of the other fcc compositions. Both the GPR and SVR achieved similar Root Mean Square Errors (RMSEs) of 22.6 and 24.8 $\frac{\text{mJ}}{\text{m} ^2}$ respectively. 

\subsection{Machine Learning: SISSO}
\label{sub:ml_sisso}
While the models developed in \cref{sub:ml_gpr_svr} are useful to navigate the chemistry-SFE space in fcc alloys, it may also be useful to generate models that are easier to evaluate and disseminate. To address this issue, the \textit{Sure Independence Screening and Sparsifying Operator} (SISSO)\cite{ouyang_sisso_2018} technique was used. In short, SISSO provides the means to generate vast feature spaces derived from primary features by combining features with predefined operators. Optimal feature subsets are selected by Sure Independent Screening \cite{Fan2008}, together with sparse operators such as Least-Absolute Shrinkage and Selection Operator (LASSO) \cite{Tibshirani1996} and $l_0$-norm regularized minimization. SISSO thus allows the generation of analytical models that are potentially more interpretable and communicable than traditional machine learning models. {Moreover, such models can also be used to extract information on feature importance or relationships between features}.

For the SISSO model, we used as primary features all 34 elemental properties and instead of applying feature selection as in the previous models, a feature space construction using the operator set: $\left [ +,-,|-|,*,^2,^3,^{0.5},ln,^{-1} \right ]$, {which is applied two times recursively, and}
resulted in a space with a final size of 391,743 features. 

\begin{figure}[htp]
    \centering
    \includegraphics[width=\textwidth]{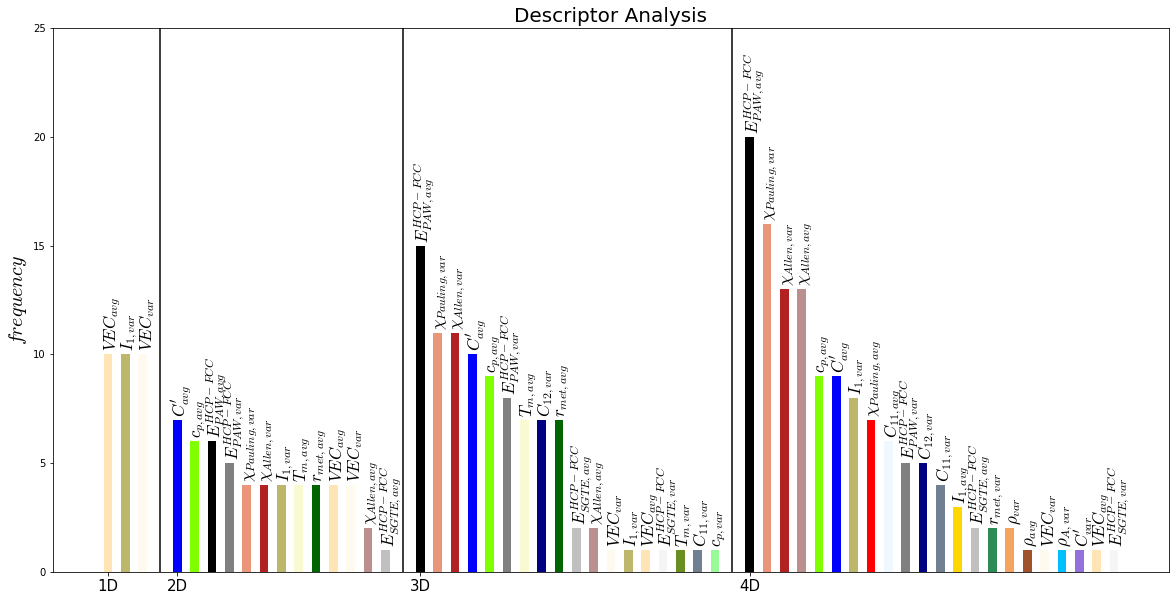}
    \caption{{The frequency of the original features present in each (1-4D) descriptor dimensions. The SISSO returns the best descriptor as shown in Eq. \ref{eq:4} and \ref{eq:5} with some of the most frequent features in dimensions 3 and 4, respectively.}}
    \label{fig:sisso_frequencies_mod}
\end{figure}

Four iterations were implemented, for each one, Sure Independence Screening (SIS) selected a subspace of 100, this selected feature subspace is cumulative and grows in steps of 100 each iteration. SIS finds the more correlated features to the SFE value at the first iteration, and to the residual of the previously chosen model. At every subspace step, {using RMSE as the error metric}, the best model is found by the Sparsifying Operator ($l_0$-norm regularized minimization) which returns an analytical linear equation of respective dimension ($1D,2D,3D,4D$). 
The analytical model obtained from a SISSO implementation in all the training data is shown in equations \ref{eq:4} and \ref{eq:5}, it achieves a coefficient of determination in the test data of 0.76 and 0.73, both of which are lower than the GPR and SVR models.  Nevertheless, the accessibility of these equations represents an easier approach to quickly estimate the SFE for an alloy of arbitrary composition, within the CoCrFeMnNiVAl chemical space. {These models appear to be consistent along all SFE calculated values. Error is comparable in the average SFE values as well as in the lower and upper boundaries for the training set. However, error in prediction has a slight positive correlation with Fe content (0.35), and a negative one with Ni (-0.29) in comparison to the rest of the concentrations: Mn (0.05), Co (-0.02), Cr (0.00), V (0.14) and Al (-0.17).} The features building this descriptor have significance for the estimation of SFE. Feature consistency is studied by analyzing 10 different analytical models from the 10-fold cross-validation---each model fit is done with different training sets initialized randomly. The frequency with which the original features are present in a coordinate of all dimensions up to 4D descriptors is shown in Figure \ref{fig:sisso_frequencies_mod}. 

\begin{figure}[htp]
    \centering
    \includegraphics[width=0.90\textwidth]{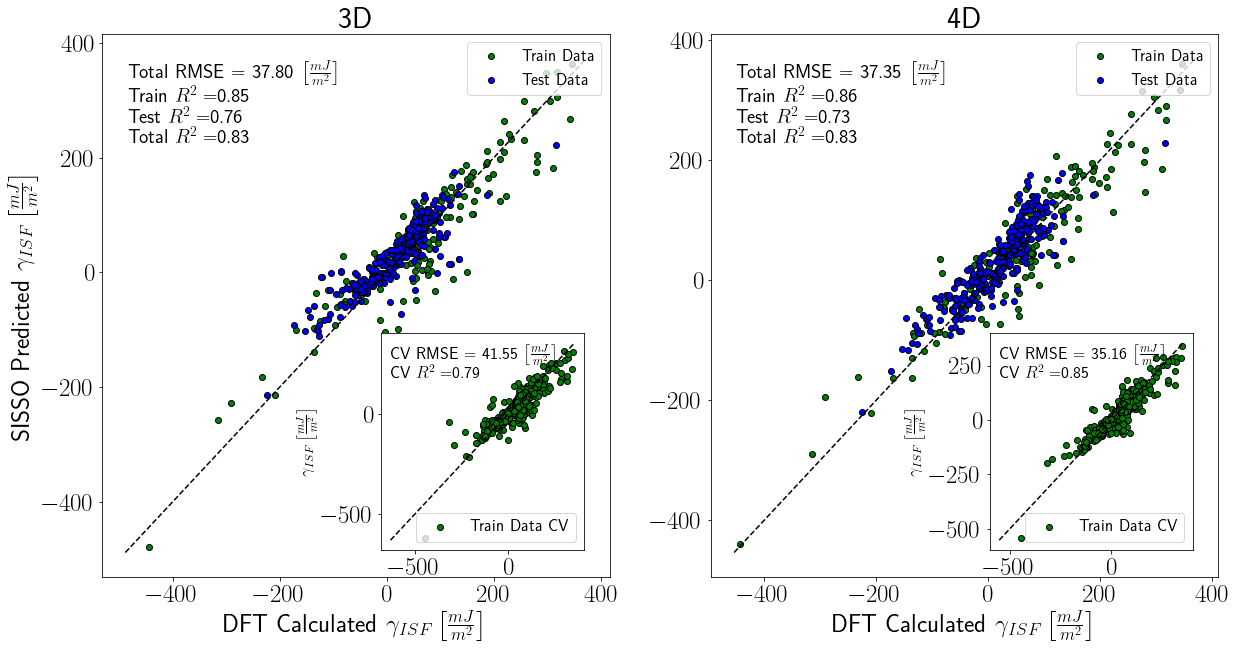}
    \caption{{The SISSO trained (a) 3D, and (b) 4D descriptors were used to train and test the medoid chosen dataset.}}
    \label{fig:best_sisso_mod}
\end{figure}

As shown in Figure \ref{fig:best_sisso_mod} the 3D descriptor has a lower test error than that of the 4D descriptor, and while the error in the training set is higher than the higher dimensional descriptor, the overall coefficient of determination is higher for the 3D descriptor. While this may indicate overfitting at a higher dimension (more complex) model, a further 10-fold cross-validation indicates that the 4D descriptors are performing better than the 3D ones on the medoid-chosen training set. That rises confidence in the final 4D descriptor for compositions further away from the clustering the chosen medoids.

SISSO returns the highest correlated feature to SFE as the first dimension, and from Figure \ref{fig:sisso_frequencies_mod}, we show that this descriptor is always one including $VEC_{var}$, $VEC_{avg}$ and $I_{1,var}$. This may indicate that while valance electrons are an important property for elastic properties, as it has been proven before\cite{music2006influence,limmer2015ab}, this property can be of more significance when damped by an energetic value such as the ionization energy. This analysis can be done for the rest of the coordinates in all the descriptors, while they are presented as individual features, their real value is brought by the combination of primary features they are in. At higher dimensions, the frequency of appearance of these features lowers and features such as the average of heat capacities, and the Pauling and Allen electronegativities variances appear in both the SISSO model and the feature selection, {in addition to some features not chosen previously like the lattice stability calculated by DFT}, and the elastic constant $C^{\prime}$. 

\begin{equation}\label{eq:4}
\begin{split}
\gamma_{ISF} & = -3.155\times 10^{3} +
1.463 \times 10^{2} I_{1,avg}^{-1}C_{p,avg}^3 
-2.972 \times 10^{-2} \left (C_{11,avg}-C_{11,var}  \right )\left (E_{PAW,avg}^\text{hcp-fcc}  \right )^3 \\ 
& +3.110 \times 10^{1} C_{var}^\prime E_{SGTE,avg}^\text{hcp-fcc}\left (\chi_{Allen,avg}-\chi_{Pauling,avg}  \right )
\end{split}
\end{equation}

\begin{equation}\label{eq:5}
\begin{split}
\gamma_{ISF} & = 4.777\times 10^{3} +
-1.609 \times 10^{2} \left (I_{1,avg}+I_{1,var}  \right )C_{p,avg}^{-1} 
+ 7.336 \times 10^{4} I_{1,var}^{-1} \chi_{Pauling,var}\rho_{A,var} \\
& +2.041 \times 10^{-1} \left (C^\prime_{avg} +C_{12,var}  \right )\left (E_{PAW,avg}^\text{hcp-fcc}  \right )^2
+1.552 \times 10^3 \left (\chi_{Allen,avg}-\chi_{Pauling,avg}  \right )E_{PAW,avg}^\text{hcp-fcc}\rho_{A,var}
\end{split}
\end{equation}


While we did not use the models represented in either equations \ref{eq:4} or \ref{eq:5} in our exploration of the fcc HEA design space, as mentioned above, such an analytical expression is quite convenient for implementation in other alloy design tools. Moreover, the physical features detected by the SISSO fitting process as being important can potentially be used to guide further exploration of the underlying physics controlling SFE in this alloy class. Satisfactorily, both SISSO and traditional ML regression models agree that the (extrapolated) lattice stability (or free energy difference) between fcc and hcp structures should be important in determining the SFE. At the same time, electronic effects (such as $VEC$) should be important in accounting for subtle alloying effects on SFE. 

\section{Overcoming the Strength-Ductility Trade-off through SFE Engineering}
\label{sec:application}


The primary advantage of a faster surrogate model is the opportunity afforded to the design process itself. {Such fast models allow one to expand the number of evaluated designs by several orders of magnitude.} If the models are fast enough, predictions---reliable up to the precision of these fast models---can be made from a dense sampling of the entire design space. In the current work, for example, 36,294 compositions have been identified as fcc at 800$^\circ$C and represent such a dense sampling. Although DFT methods were used to calculate the SFE of only 498 of these compositions, the regression models trained on this data can be used to predict the SFE of the remaining compositions and evaluate their value as potential designs. A prevailing theme among the current state-of-the-art in HEA design is the maximization of both strength and ductility. These quantities are complex properties {that depend on} a variety of mechanisms at multiple scales, but both can be approximated by more achievable metrics. 

As discussed above, the stacking-fault energy ($\gamma_\text{ISF}$) can be used to predict, or at least infer, the likely deformation mechanisms and therefore contributors to the ductility of an fcc alloy. An SFE value of around 20 $\sfrac{\text{mJ}}{\text{m} ^2}$ has been shown to be the most likely to produce favorable deformation mechanisms like TRIP and TWIP~\cite{chaudhary2017data}. This SFE value has also been estimated in HEAs that exhibit TWIP/TRIP behavior\cite{su2019hierarchical,wu2021microstructure}. The closer alloys are to this target value, the more likely they are to activate additional plasticity {mechanisms} and increase {their} ductility, {potentially} overcoming the strength-ductility trade-off. While SFE plays a role in the plastic deformation mechanisms of a given alloy, its baseline resistance to deformation is ultimately controlled by its intrinsic strength. The investigation of the strength-ductility trade-off would then require {the additional use} of a fast-acting model capable of predicting the strength of a given alloy as a function of composition.

Recently, Oh et al. introduced a model {for the } solid-solution strengthening in fcc HEAs and verified its accuracy for the Cantor alloy and related systems \cite{oh2019engineering}. Their model assumes that the intrinsic strength in fcc transition metal HEAs depends on heterogeneous internal "pressure" fields ultimately associated to the heterogeneous charge transfer between dissimilar constituents in the alloy. While {the analysis} of such charge transfer contribution to strength would require highly sophisticated quantum mechanical calculations based on DFT, Oh et al. found that they could correlate the heterogeneous local charge transfer with the variance in the electronegativity of the alloy constituents. As shown in Equation~\ref{eq:strength_oh}, the modeled strength [MPa] depends only on the composition-weighted variance in Allen electronegativity, which is trivial to calculate for an arbitrary composition and was {(incidentally} already included in the surrogate model training as a property-based feature. 

\begin{equation}
    \sigma_{\text{SS}} = 4293 \chi_{\text{Allen, var}} + 84~[\text{MPa}]
\label{eq:strength_oh}
\end{equation}

Although strength and ductility are commonly contradictory properties in complex alloys, solid-solution strength and stacking fault energy can be used to visualize this trade-off and identify alloys that push the boundaries of the state-of-the-art intrinsic strength and ductility. In this case, we use the distance from SFE~20 $\sfrac{\text{mJ}}{\text{m} ^2}$ as the surrogate for the plasticity-enhancing twinning deformation mechanism. We would like to note that other SFE targets could potentially be used, depending on the {desired plastic deformation mechanism}. Since our trained model(s) are capable of predicting SFEs over the entire range (from negative to positive values), possible targets include TWIP, TRIP, TWIP/TRIP deformation mechanisms. We note that even the micro-faulting mechanism due to negative intrinsic SFE recently reported by Wei and Tasan~\cite{wei2020deformation} {can potentially be a target for our multi-objective optimization}.

\begin{figure}[H]
    \centering
    \includegraphics[width=\textwidth]{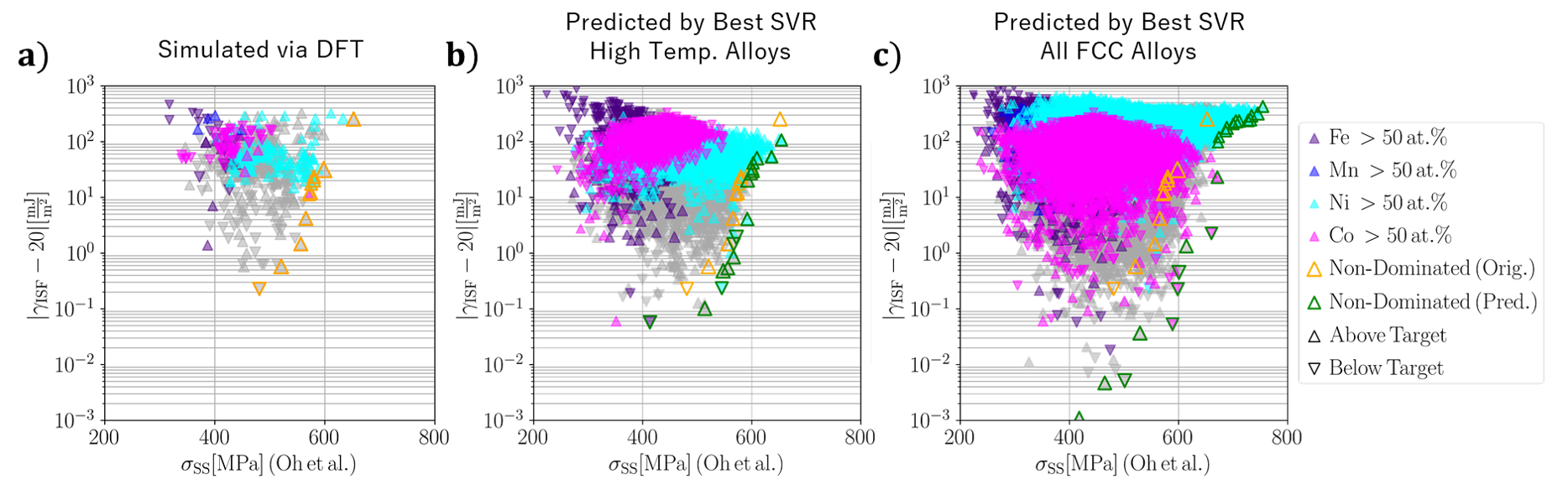}
    \caption{a) Alloys plotted in a multi-objective space between solid-solution strength (Eqn.~\ref{eq:strength_oh}) and the distance to an SFE target of 20 $\sfrac{\text{mJ}}{\text{m} ^2}$ (logarithmic axis) for alloys simulated with DFT. b-c) The same objective space as predicted by the best support vector regressor for b) high temperature alloys that satisfied the Scheil conditions in Fig.~\ref{fig:calphad_figures} and c) all alloys predicted to be FCC at 800 $^\circ$C. Non-dominated points on the Pareto frontier are highlighted in each plot.}
    \label{fig:pareto_plots}
\end{figure}

The trade-off surface of a multi-objective design problem is known as a Pareto frontier. Points along this frontier are referred to as \textit{non-dominated}, meaning that no other point offers improvement {in at least one direction/objective}. Figure~\ref{fig:pareto_plots}a shows the calculated SFE (in reference to a target of 20 $\sfrac{\text{mJ}}{\text{m} ^2}$) plotted against the solid-solution strength, using the model by Oh \emph{et. al} of the original 498 compositions simulated via DFT. The lower right surface of this plot represents the Pareto frontier of the most promising alloys: those nearest the desired SFE and with the highest predicted strength. Eleven compositions in the original dataset are non-dominated, meaning no other composition has both a higher strength and an SFE closer to the target. Of these, four have a majority constituent element (three Ni and one Cr) and eight have Ni concentrations greater than 40 at.\%.

To expand the Pareto frontier of potential alloys, the best support vector regressor that was trained on the 498 DFT calculations was used to predict the SFE of the 5,739 alloys suitable for high temperature applications (i.e. those that satisfied the Scheil constraints in Fig.~\ref{fig:calphad_figures}). The predicted SFE (in reference to the target) and solid solution strength are visualized in Fig.~\ref{fig:pareto_plots}b. Of these \emph{potential} high temperature alloys, 235 are not dominated by the original DFT model, meaning the SVR model of SFE was able to identify over 200 alloys that are more likely to push the boundaries of {the} known property space. Furthermore, the SVR model was applied to all 36,294 alloys predicted to be FCC at 800 $^\circ$C, which provides a thorough representation of the FCC region in the CoCrFeMnNiV-Al space. Figure~\ref{fig:pareto_plots}c visualizes the strength-ductility property space for this set. By using the SVR on this expanded set, an additional 2089 alloys were discovered that are not dominated by the original DFT data. The new Pareto front consists of 22 non-dominated compositions. As shown in Figure~\ref{fig:pareto_plots}c, the predicted strength of this new expanded front is 100 MPa higher than the original front in regions close to the target SFE. The compositions on or near this front represent candidate alloys that are most likely to produce the desired deformation mechanisms, increasing their ductility, while also exhibiting high intrinsic strength. A list of such alloys, as well as some predicted properties, can be found in the Supplementary Material. 

Table~\ref{table:example_compositions} lists the ten compositions with the highest predicted solid solution strength (Eqn.~\ref{eq:strength_oh}) and predicted stacking fault energies between 15 and 25 mJ/m$^2$ in the high-temperature set (solidus temperature greater than 1600 K and solidification range below 100 K). These alloys and other alloys in the updated Pareto frontier are a significant result as these new potential targets for further investigation constitute true predictions resulting from the combination of two reasonably accurate fast-acting models. Moreover, these alloys correspond to compositions that already pass a number of stringent but reasonable performance constraints that are ultimately unavoidable if one is interested in the design of optimal \textbf{and} feasible fcc HEAs. While the present analysis has been limited to a very narrow target for the value of SFE (20 $\sfrac{\text{mJ}}{\text{m} ^2}$), we would like to point out that a similar analysis could be carried out to identify alloys with SFEs arbitrarily close to any other value, from negative to positive, in turn corresponding to different potential dominant plastic deformation mechanisms. Identification of such alloys could enable a more efficient exploration of the fcc HEA space. More importantly, designing alloys {with a specific SFE} that also have a reasonable probability of being synthesizable could enable a more systematic study of the 'plastic deformation landscape' in fcc HEAs.

\begin{table*}[h]
\caption{Sample of Compositions Suggested for Further Investigation}
\label{table:example_compositions}
\begin{center}
\begin{tabular}{| c | c | c | c | c | c | c | c | c | c |}
\hline

Co & Cr & Fe & Mn & Ni & V & Al & \multirow{2}{*}{$S_\text{conf.}$} & Predicted & Predicted \\

[at.\%] & [at.\%] & [at.\%] & [at.\%] & [at.\%] & [at.\%] & [at.\%] & & $\sigma_\text{SS}$ [MPa] & $\gamma_\text{ISF}$ [mJ/m$^2$] \\

\hline
\rowcolor[HTML]{D9D9D9} 
8.7\%  & 0.0\%  & 24.0\% & 0.2\% & 51.2\% & 13.2\% & 2.7\% & 1.28 & 592 & 24.1 \\
17.5\% & 9.1\%  & 16.5\% & 0.5\% & 45.0\% & 9.7\%  & 1.6\% & 1.50 & 572 & 18.1 \\
\rowcolor[HTML]{D9D9D9} 
20.3\% & 27.3\% & 3.0\%  & 0.5\% & 43.4\% & 4.2\%  & 1.2\% & 1.36 & 567 & 18.6 \\
12.4\% & 1.7\%  & 23.7\% & 0.7\% & 48.3\% & 11.6\% & 1.5\% & 1.37 & 567 & 20.8 \\
\rowcolor[HTML]{D9D9D9} 
26.4\% & 18.6\% & 2.7\%  & 0.3\% & 41.9\% & 4.7\%  & 5.4\% & 1.44 & 567 & 24.8 \\
16.4\% & 8.8\%  & 19.0\% & 0.4\% & 43.4\% & 8.9\%  & 3.2\% & 1.53 & 566 & 21.0 \\
\rowcolor[HTML]{D9D9D9} 
8.8\%  & 0.5\%  & 19.5\% & 0.1\% & 58.7\% & 11.9\% & 0.4\% & 1.16 & 566 & 22.4 \\
16.1\% & 4.9\%  & 22.3\% & 0.2\% & 42.9\% & 9.7\%  & 3.9\% & 1.50 & 565 & 16.7 \\
\rowcolor[HTML]{D9D9D9} 
22.7\% & 18.0\% & 4.6\%  & 0.2\% & 46.2\% & 5.5\%  & 2.8\% & 1.42 & 562 & 17.0 \\
14.7\% & 1.5\%  & 21.4\% & 0.4\% & 48.5\% & 10.4\% & 3.2\% & 1.39 & 561 & 24.0 \\
\hline
\end{tabular}
\end{center}
\end{table*}

\section*{Summary and Conclusions}

Stacking Fault Energy (SFE) is an intrinsic alloy feature that is tightly coupled to the mode of plastic deformation of fcc alloys. It is by activating plastic deformation mechanisms in addition to slip that {it is possible} to overcome the strength-ductility trade-off  {that constrains the performance space in structural materials}.  Here we focused on developing tools for a more systematic investigation of the composition dependence of the SFE in fcc high-entropy alloys (HEAs), with the major theme being "design for specific plastic deformation mechanisms". A major obstacle to the use of SFE in alloy performance optimization is that its estimation is exceedingly computationally costly in most theoretical approaches. Estimating SFE using experimental means is even more challenging, {and in some cases limited by the bias in the models ultimately used to interpret experiments and infer the value for this intrinsic material property.} The lack of accurate means to estimate SFE of arbitrary compositions implies that this very important alloying indicator is mostly used to rationalize {the observed} behavior of alloys under mechanical deformation, rather than as a 'design metric' amenable to optimization. 

In this work, we seek to bridge the existing gap in the scientific community focused on HEA development by investigating the dependence of chemistry on the intrinsic SFE in the fcc CoCrFeMnNiV-Al HEA space.  Given the vast chemical space, we filtered much of the available space by first carrying out a CALPHAD-based exhaustive search, focused on identifying feasible regions in this six-dimensional space corresponding to alloys with some probability of being synthesizable and with reasonable phase stability characteristics. We then employed reliable \textit{first-principles} DFT-based estimates of SFE within the axial next-nearest-neighbor Ising (ANNNI) formalism, and confirmed these estimates by direct SFE calculations. This approach provides accurate predictions for SFE (as compared with scarce experimental inferences) without the considerable uncertainty involved in explicit calculations of the Generalized Stacking Fault Energy surface. The DFT-estimated SFEs constitute, to the best of our knowledge, the largest dataset for SFE in fcc HEAs. From this dataset, we developed an accurate and easy to evaluate machine-learning (ML) models, including a fully analytical expression for SFE as a function of alloy features developed through the SISSO framework.

These ML models in materials science are not useful without the opportunity to deploy them to optimize a materials (design) space. We thus proceeded to combine the SFE models with equally easy-to-deploy physics-based models for intrinsic strength {to explore the Pareto front of the strength-ductility trade-off}. Our analysis shows that the deployment of models for SFE and for strength resulted in candidate alloys about 100 MPa stronger than what could have been inferred by simply focusing on  alloys investigated explicitly through DFT methods, illustrating the usefulness of the ML-assisted approach. The examination of the Pareto front also shows that there is no strong correlation between strength and SFE. This finding is significant as it implies that it is potentially possible to design \emph{feasible} fcc HEAs with specific intrinsic strength levels and dominant plastic deformation mechanisms. This approach also provides more freedom in the design of fcc HEAs with additional performance requirements, other than those considered here.

\section*{Associated Content}
DFT-calculated stacking fault energies for 498 compositions and Python source code for property-based feature generation and regression model training and testing is available at CodeOcean: \url{https://codeocean.com/capsule/2312007/tree}

\section*{Acknowledgements}
GV, KY and RA acknowledge the support of QNRF under Project No. NPRP11S-1203-170056. TK(irk) acknowledges the support of NSF through Grant No. NSF-DGE-1545403. TK(han) initiated this work during an Online REU (O-REU) research opportunity at Texas A\&M University, supported in part by NSF through Grants No. NSF-DEMS-1663130 and NSF-DMREF-1729350. DFT calculations were carried out in part at the Texas A\&M High-Performance Research Computing (HPRC) Facility. Work at Ames Laboratory was funded by the U.S. Department of Energy (DOE), Office of Science, Basic Energy Sciences, Materials Science and Engineering Division. Ames Laboratory is operated for the U.S. DOE by Iowa State University under Contract No. DE-AC02-07CH11358. TK(han) carried out the feature selection analysis and GPR and SVR ML training. PS carried out SFE and electronic-structure calculations using DFT for the fcc alloys. AVS helped with DFT-KKR-CPA code. PS and DDJ analyzed the DFT data. TK(irk) supervised TK(han) and carried out the CALPHAD-based exploration of the fcc HEA space. GV did the SISSO analysis. RA designed and supervised the project. All authors contributed to the writing of the manuscript. RA, TK and PS carried out the final revisions/edits of the manuscript.

\section*{Appendix}
\subsection*{Appendix A: Visualization of Top Property-based Features}

\begin{figure}[H]
    \centering
    \includegraphics[width=\textwidth]{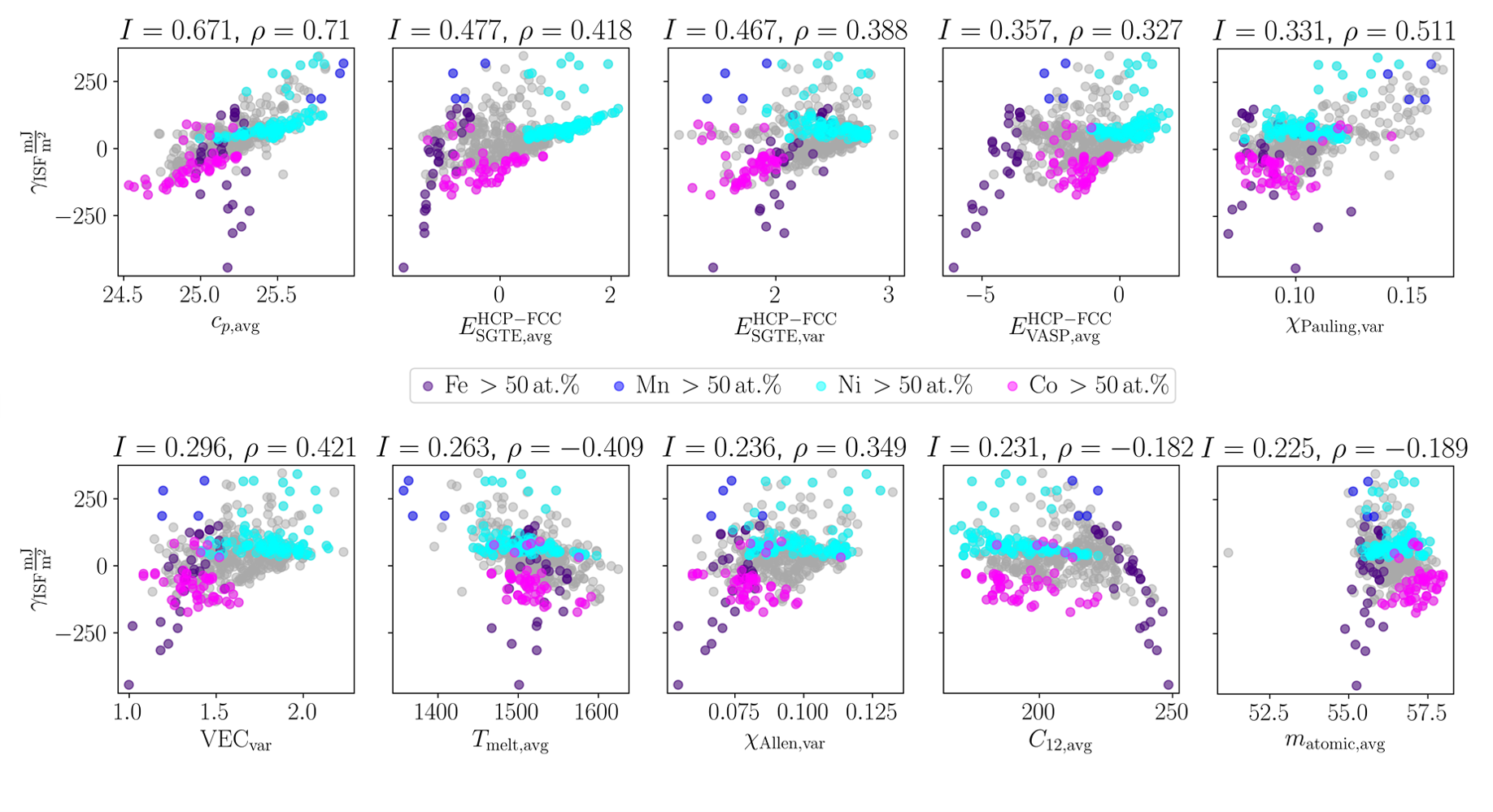}
    \caption{The ten property-based features with the highest mutual information ($I$) with stacking fault energy plotted against stacking fault energy. Above each plot is the corresponding mutual information and Pearson correlation coefficient ($\rho$) between the respective feature and stacking fault energy.}
    \label{fig:feature_scatter_plots}
\end{figure}

Figure~\ref{fig:feature_scatter_plots} displays the property-based features with the highest mutual information with stacking fault energy. Most of the top performing features have positive correlations. Note that the outlying datapoint in the atomic mass plot is the equiatomic composition (Co$_{14.3}$Cr$_{14.3}$Fe$_{14.3}$Mn$_{14.3}$Ni$_{14.3}$V$_{14.3}$Al$_{14.3}$ [at.\%]) because it has higher aluminum content than any other datapoint.

\subsection*{Appendix B: Exploration of Alternative Feature Spaces}

While the 34 property-based features that were originally selected cover a wide range of physical phenomena, the potential set of material properties is much larger. Wang et al. \cite{Wang2020bestpractices} have presented a method for quickly generating so-called Composition-Based Feature Vectors (CBFVs) from large databases of material properties. To compare the performance of other property-based features, Wang et al.'s approach to generating CBFVs was used to generate 44 stoichiometrically averaged properties from the Oliynyk \cite{oliynyk_high-throughput_2016} and from the Magpie \cite{ward_general-purpose_2016} {feature sets}. 

\begin{figure}[H]
    \centering
    \includegraphics[width=\textwidth]{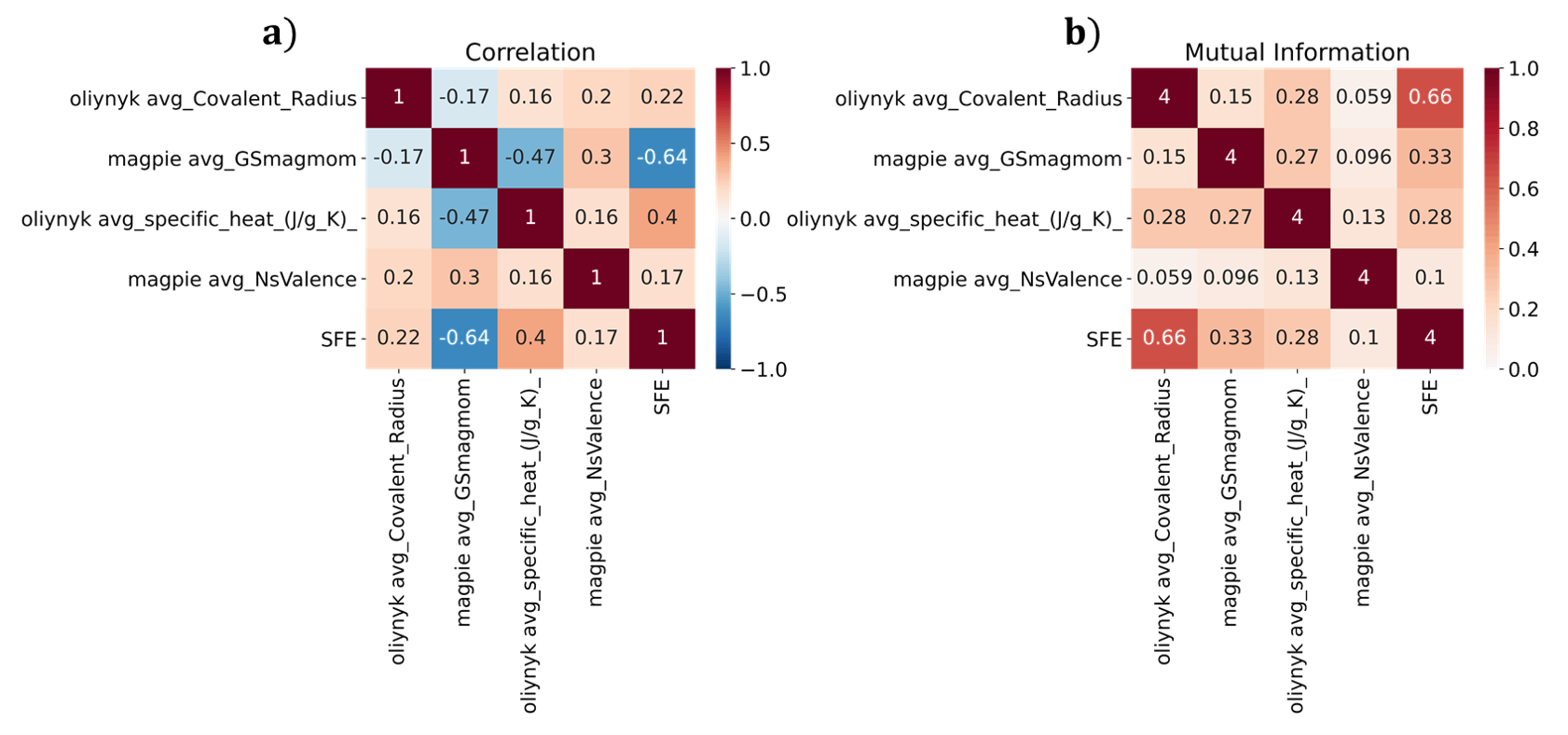}
    \caption{Matrices displaying pairwise a) Pearson correlation coefficients and b) mutual information between the four features selected from the Magpie \cite{ward_general-purpose_2016} and Oliynyk \cite{oliynyk_high-throughput_2016} datasets and stacking fault energy.}
    \label{fig:expanded_feature_matrices}
\end{figure}

The same feature selection process employed to {down-select} the original 34 features was performed for the new set of 66 features. By sorting features by descending mutual information with SFE and ensuring selected features had pairwise correlations less than 0.5 with each other, a final set of four descriptive yet independent features were chosen. The four selected features were the average covalent radius and average specific heat from the Oliynyk dataset and the average magnetic moment and average number of valence electrons from the Magpie dataset. Pairwise mutual information and correlation matrices for the selected features are shown in Fig.~\ref{fig:expanded_feature_matrices}. Two of these properties were considered in the original feature set: specific heat and the number of valence electrons. The average specific heat was also selected in the original process and had the highest observed mutual information with SFE (Fig.~\ref{fig:feature_scatter_plots}). We note that while not selected for the GPRs or SVRs, the number of valence electrons was used as a feature in many of the {single-term SISSO models}, as seen in Fig.~\ref{fig:sisso_frequencies_mod}. 

\begin{figure}[H]
    \centering
    \includegraphics[width=\textwidth]{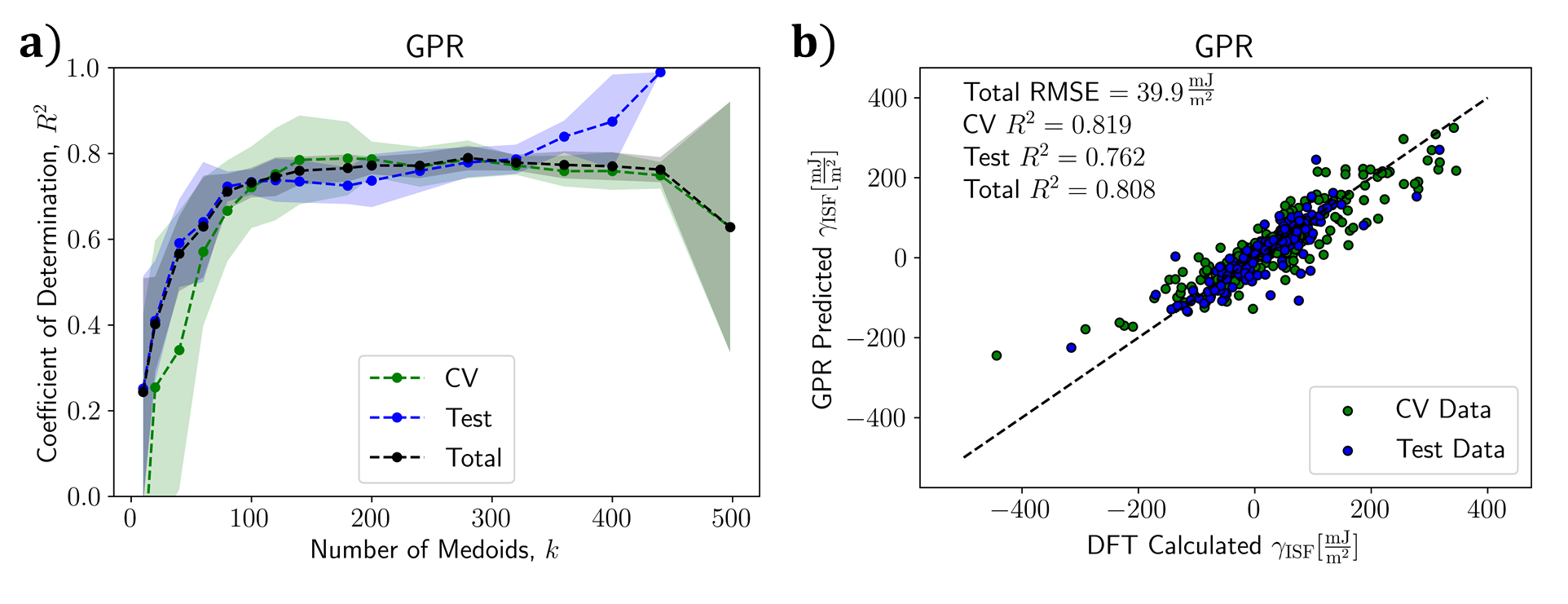}
    \caption{a) Coefficient of Determination ($R^2$) of each trained Gaussian process model {(Mat\'{e}rn kernel)} versus the number of medoids selected for training ($k$). b) Predicted versus actual stacking fault energy for the best Gaussian process regressors. Predicted values for the training data were obtained via cross-validation (CV Data). The model was also tested against data points that were not chosen as medoids (Test Data).}
    \label{fig:expanded_gpr}
\end{figure}

The four chosen features were used to train Gaussian process regressors {(Mat\'{e}rn kernel)} in a similar procedure to the {one in which the} original feature set {was used}. The accuracy of the GPRs as a {function of training set size} and {a parity plot for the best GPR are shown in Figs.~\ref{fig:expanded_gpr}a and b, respectively. The new GPRs generally have lower accuracies than those trained on the original feature set. Also  Fig.~\ref{fig:expanded_gpr}a shows a slight decrease in accuracy as the training set size reaches a maximum, unlike the models in ~\ref{fig:models_vs_k}a. The best model trained on the new feature set (Fig.~\ref{fig:expanded_gpr}b) was trained with 280 data points and has a Total $R^2$ of 0.808 compared to a Total $R^2$ of 0.939 for the original best GPR (Fig.~\ref{fig:best_models}a).} The general difference in performance between the feature sets can be explained by the exclusion of properties that are particularly predictive of SFE, like the energy differences between hcp and fcc lattices, or stochiometric property variances, which are often independent from stochiometric property averages.

\subsection*{Appendix C: Influence of Kernel Choice on Regressor Accuracy}

\begin{figure}[H]
    \centering
    \includegraphics[width=\textwidth]{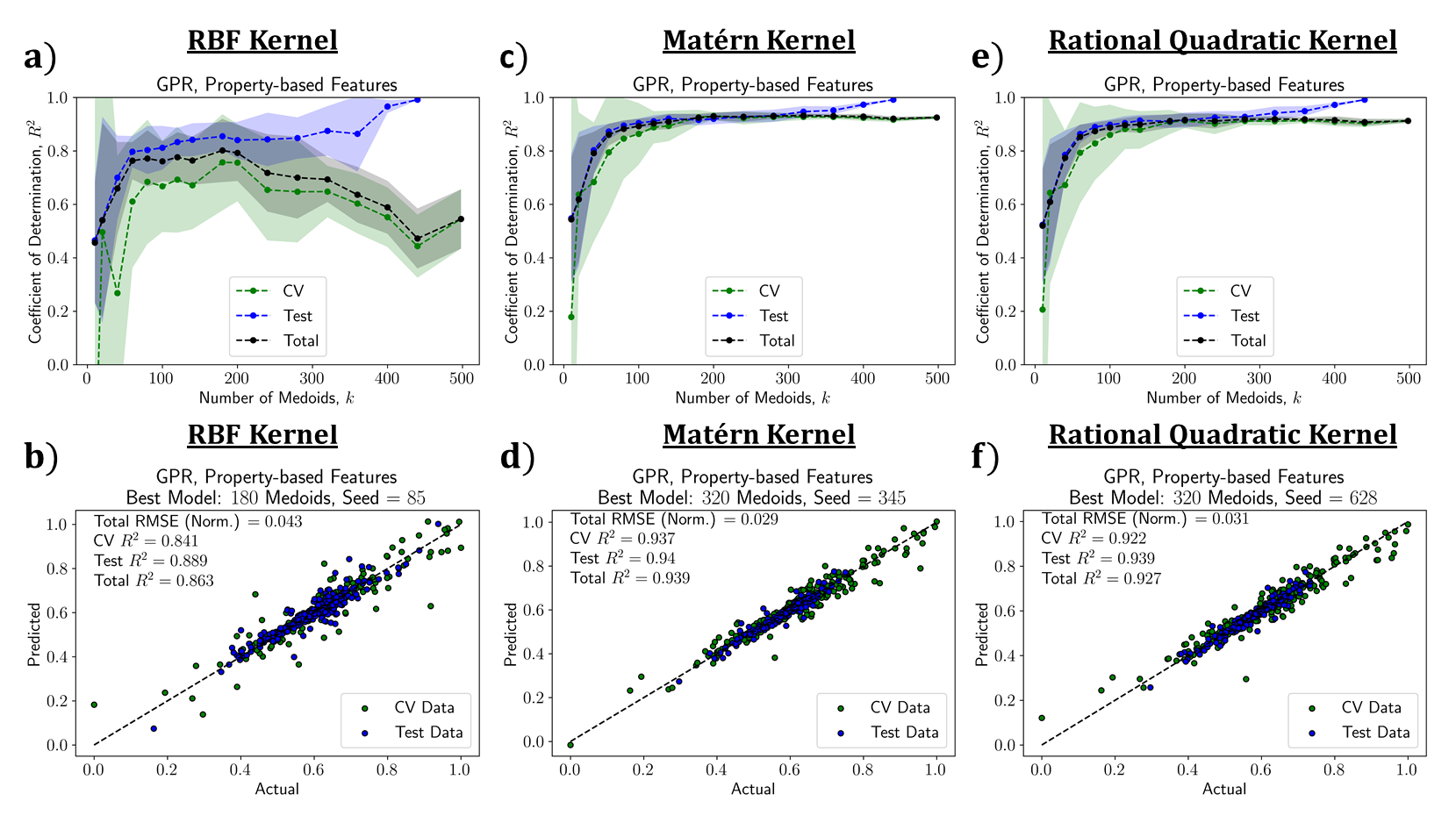}
    \caption{{Coefficient of determination for ten random runs as a function of training size as well as predicted versus actual data for Gaussian Processor Regressors (GPRs) trained with (a \& b) RBF kernels, (c \& d) Mat\'{e}rn kernels, (e \& f) Rational Quadratic kernels.}}
    \label{fig:gpr_all_kernels}
\end{figure}

{Figure~\ref{fig:gpr_all_kernels} displays the results of GPRs trained with Radial Basis Function (RRF) kernels,  Mat\'{e}rn kernels, and Rational Quadratic kernels. The RBF models exhibits significantly worse performance than either of the other kernels. The accuracy of the GPRs with RBF kernels reached a maximum near 200 medoids, but then the total $R^2$ decreased to below $0.6$ as the size of the training set increases. This decrease is largely due to poor predictions on the cross-validated set, which arise from the bias in the training data. As the training set grows to contain most of the calculated data, the GPRs become overfit to the HT compositions during cross-validation. This overfitting results in poor prediction of the other fcc data in the training set, but performance on the test set, which contains mostly HT data, remains high.}

{The RBF kernel performs worse because it generates smoother functions that are more prone to sample bias. The Mat\'{e}rn kernel overcomes sample bias by varying the smoothness with an additional parameter. Similarly, the Rational Quadratic kernel also avoids the influence of sample bias by combing RBF kernels of various scales. Caution should be used when fitting smooth functions to potentially biased sample sets. Such functions can be overfit in oversampled areas if hyperparameter objectives consider aggregate error metrics that neglect large residuals in a minority of samples.}

\begin{figure}[H]
    \centering
    \includegraphics[width=\textwidth]{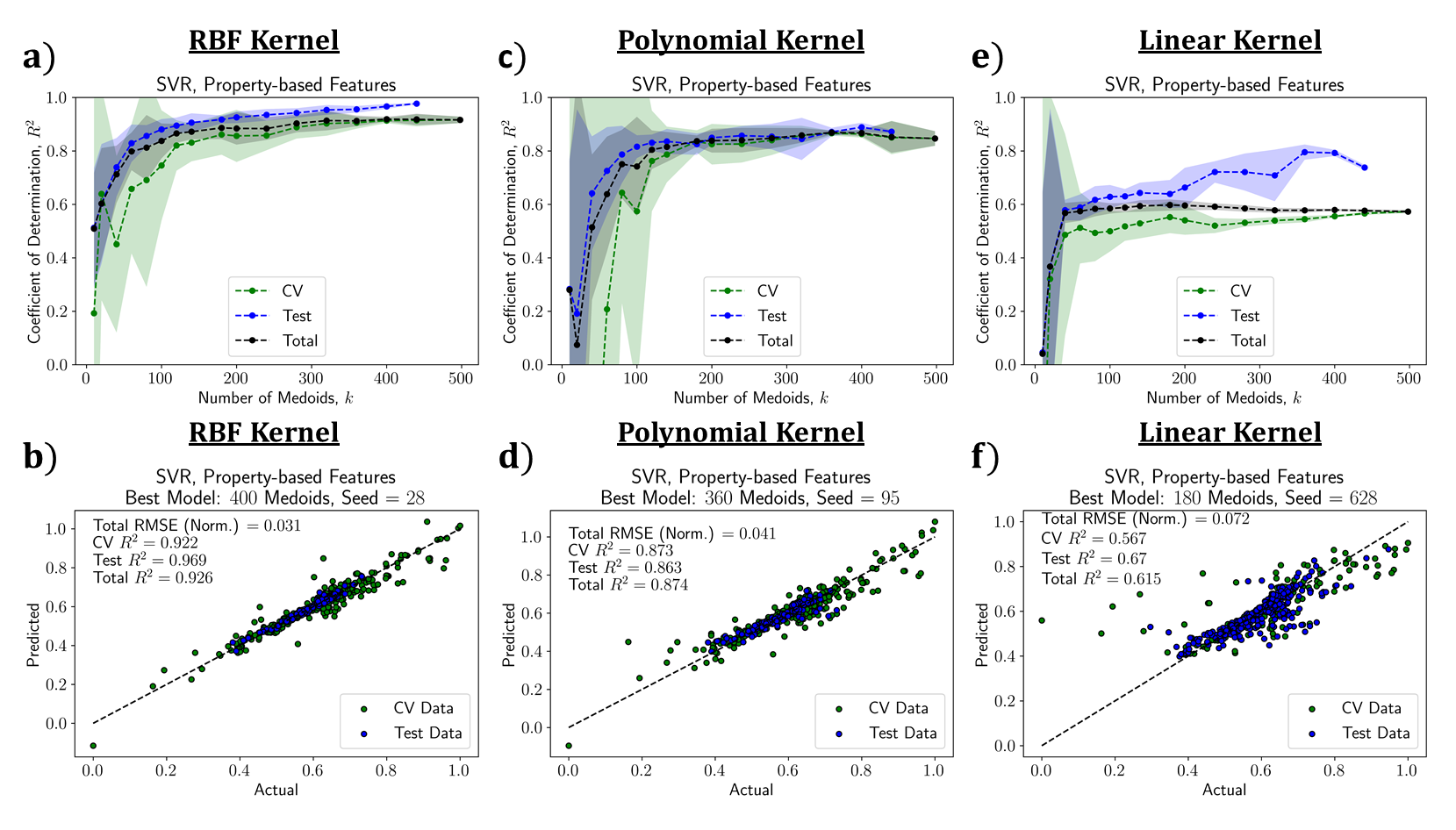}
    \caption{{Coefficient of determination for ten random runs as a function of training size as well as predicted versus actual data for Support Vector Regressors (SVRs) trained with (a \& b) RBF kernels, (c \& d) Polynomial kernels, (e \& f) Linear kernels.}}
    \label{fig:svr_all_kernels}
\end{figure}

{Figure~\ref{fig:svr_all_kernels} displays the results of SVRs trained with Radial Basis Function (RRF) kernels,  polynomial kernels, and linear kernels. The RBF kernel showed the best performance while the polynomial kernel was slightly worse and  the linear was significantly worse. Unlike the GPR with the RBF kernel, none of the SVRs seemed susceptible to sample bias (i.e. no decrease in accruacy as the training size increased). SVRs ensure the residuals of all samples are below a uniform threshold, which seems to help mitigate sample bias in the training set. The decreased performance of the polynomial and linear kernels can likely be attributed to the inability of simpler models to capture complexity in the response.}

\end{document}